%% file: exir-sigconf-authordraft.tex
\documentclass[sigconf,natbib=true,anonymous=false,nonacm]{acmart}
\renewcommand\footnotetextcopyrightpermission[1]{}

\AtBeginDocument{%
  }

\usepackage{orcidlink}
\usepackage{multirow}
\usepackage{todonotes}
\usepackage{microtype}
\usepackage{subcaption}
\usepackage{tcolorbox}
\tcbuselibrary{skins}
\begin{document}

\title{ExpertLens: Visualizing Embedding Spaces for Post-Hoc Explainability in MoE Enhanced Retrievers}



\author{Effrosyni Sokli\orcidlink{0009-0003-5388-2385}}
\email{effrosyni.sokli@unimib.it}
\affiliation{%
  \institution{Department of Informatics, Systems and Communication, DISCo \\ University of Milano-Bicocca}
  \city{Milan}
  \country{Italy}
}

\author{Isaac Roberts\orcidlink{0009-0005-3875-729X}}
\email{iroberts@techfak.uni-bielefeld.de}
\affiliation{%
  \institution{CITEC - Center for Cognitive Interaction Technology \\ Bielefeld University}
  \city{Bielefeld}
  \country{Germany}
}

\author{Alexander Schulz\orcidlink{0000-0002-0739-612X}}
\email{aschulz@techfak.uni-bielefeld.de}
\affiliation{%
  \institution{CITEC - Center for Cognitive Interaction Technology \\ Bielefeld University}
  \city{Bielefeld}
  \country{Germany}
}

\author{Barbara Hammer\orcidlink{0000-0002-0935-5591}}
\email{bhammer@techfak.uni-bielefeld.de}
\affiliation{%
  \institution{CITEC - Center for Cognitive Interaction Technology \\ Bielefeld University}
  \city{Bielefeld}
  \country{Germany}
}

\author{Gabriella Pasi\orcidlink{0000-0002-6080-8170}}
\email{gabriella.pasi@unimib.it}
\affiliation{%
  \institution{Department of Informatics, Systems and Communication, DISCo \\ University of Milano-Bicocca}
  \city{Milan}
  \country{Italy}
}

\renewcommand{\shortauthors}{Sokli et al.}

\begin{abstract}
Neural models, including dense retrievers, have been widely adopted in Information Retrieval (IR), often delivering state-of-the-art performance.
Despite their effectiveness, these models operate as black boxes, limiting the interpretability of their ranking decisions.
Existing post-hoc explainability methods for neural rankers primarily focus on feature-level attributions, which can be insufficient to capture the complexity of learned embedding spaces.
In this work, we propose ExpertLens, a post-hoc explainability framework for Mixture-of-Experts (MoE)-enhanced dense retrievers that shifts focus from local scalar feature importance to representation-level global interpretability.
ExpertLens leverages discriminative embedding space visualizations jointly with automatically extracted Concept Activation Vectors to reveal how expert routing drives embedding space formulation and retrieval effectiveness.
Experiments across five IR benchmarks and two MoE-enhanced dense retrievers show that expert routing consistently improves embedding space structure, positioning queries and their relevant documents into better-defined geometric neighborhoods.
Analysis of expert subspaces further reveals general-purpose dominant experts, along with minority experts exhibiting distinct linguistic specialization, with subspaces arranged according to multi-semantic concept similarity. 
Our code is available at: \url{https://anonymous.4open.science/r/ExpertLens}.
\end{abstract}




\maketitle

\section{Introduction}
The recent and increasing use of neural networks in Information Retrieval (IR) has reshaped how retrieval systems estimate relevance, moving from lexical matching toward semantic representation learning.
Various IR approaches (e.g., learning-to-rank algorithms) rely on selected features to learn scoring functions offering partial interpretability to the model's ranking decisions.
Neural sparse retrievers assign learned lexical weights to query and document tokens, enabling efficient inverted-index retrieval while providing partial interpretability through token-level weight inspection~\cite{DBLP:conf/sigir/FormalPC21}.
However, their reliance on explicit lexical signals can limit their ability to capture deep semantic relationships.
In contrast, dense retrievers encode queries and documents into high-dimensional embeddings, capturing semantic relationships beyond explicit lexical matching~\cite{DBLP:journals/ftir/MitraC18}.
While dense retrievers have demonstrated strong performance across many IR benchmarks, they behave as black boxes, making their ranking decisions difficult to interpret.
Understanding how their internal representations structure the embedding space and drive retrieval decisions remains an open challenge, and a necessary step toward interpretable neural IR systems.

More recently, modern neural rankers and dense retrievers leverage Mixture-of-Experts (MoE)-based architectures~\cite{DBLP:conf/icml/HoulsbyGJMLGAG19,DBLP:journals/neco/JacobsJNH91} to tailor general-purpose Language Models (LMs) to IR tasks.
Specifically, MoE dynamically route inputs through specialized experts, allowing models to scale efficiently while retaining task-specific expertise~\cite{DBLP:conf/iclr/LiSYWRCZ023,DBLP:conf/iclr/ShenHZ0LWCZFCVW24}.
In IR, MoE-enhanced retrievers have demonstrated promising gains in retrieval effectiveness, task generalization, and domain adaptation~\cite{DBLP:conf/nlpcc/DaiJZLSC23,DBLP:journals/corr/abs-2304-10195,DBLP:journals/tois/GuoCBFCZC25}, establishing MoE as a prominent paradigm in neural IR.
At the same time, the high-dimensional representations inherent to these models challenge the ability of traditional feature attribution methods to provide adequate explanations.
Existing post-hoc explainability approaches for neural rankers largely operate at the feature level and have been adapted from classification to search tasks, producing attributions that estimate the contribution of individual input features to ranking scores~\cite{DBLP:conf/kdd/Ribeiro0G16,DBLP:conf/nips/LundbergL17,DBLP:conf/sigir/HeussRA25}.
Understanding how ranking behavior and textual representations evolve under expert specialization requires moving beyond scalar feature importance towards interpretable structure within embedding spaces.

In this paper, we introduce ExpertLens, a global, representation-level post-hoc explainability framework~\cite{opitz2025interpretable} tailored to MoE-enhanced dense retrievers.
Instead of assigning attribution scores to individual input features, ExpertLens examines how expert routing affects the geometry of the embedding space and organizes document representations structurally.
We leverage discriminative embedding space visualizations derived from DeepView~\cite{DBLP:conf/ijcai/0001HH20}, a Discriminative Dimensionality Reduction (DiDi) framework that projects high-dimensional representations into interpretable, discriminative views.
We evaluate ExpertLens on five IR benchmarks and two MoE-enhanced dense retrievers with contrasting gating mechanisms, isolating how expert routing shapes the induced embedding spaces.
Our main contributions are:
\begin{enumerate}
    \item We leverage DeepView to analyze the embedding spaces produced by MoE-enhanced dense retrievers, providing interpretable insights into expert routing and embedding space structure across domains and IR tasks.
    \item We conduct qualitative and quantitative analyses of expert-induced subspaces, characterizing expert specialization in terms of linguistic properties and semantic themes.
    \item We use automatically extracted Concept Activation Vectors (CAVs)~\cite{kim2018interpretabilityfeatureattributionquantitative} to associate semantic attributes with the underlying structure of the embedding space.
\end{enumerate}
ExpertLens provides explainable insights into both the embedding space structure and the specialization of experts.
Our approach complements existing feature-level attribution methods to advance the post-hoc global explainability of MoE-enhanced dense retrievers.

\begin{figure*}[t]
  \centering
  \includegraphics[width=\linewidth]{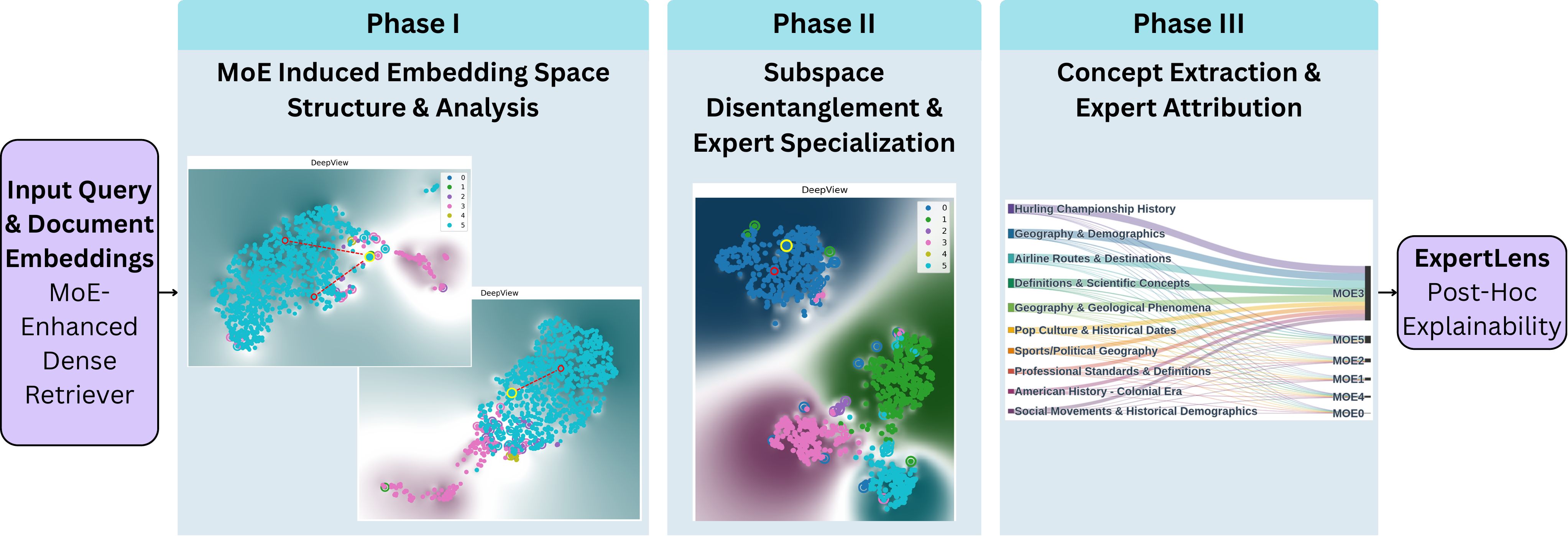}
  \caption{The ExpertLens framework. \textbf{Phase~I:} Visualization of the MoE induced embedding space through discriminative DeepView projections for structure analysis. \textbf{Phase~II:} Examination of the partitioning of the embedding space into subspaces and association with expert specialization. \textbf{Phase~III:} CAVs are automatically extracted and mapped to individual experts.}
  \label{fig:framework}
\end{figure*}
\section{Related Work and Background}
Explainability in IR (ExIR) has traditionally centered on feature attribution methods, with foundational techniques such as SHAP~\cite{DBLP:conf/nips/LundbergL17} and LIME~\cite{DBLP:conf/kdd/Ribeiro0G16} adapted to ranking tasks to quantify the contribution of individual query-document features to relevance predictions.
LIRME~\cite{DBLP:conf/sigir/VermaG19} extended this paradigm with local, model-agnostic interpretability for neural rankers, and RankingSHAP~\cite{DBLP:conf/sigir/HeussRA25} formalized listwise Shapley value estimation that also accounts for interdependencies among ranked documents.
The growing adoption of neural models in IR, especially MoE architectures \cite{DBLP:conf/nlpcc/DaiJZLSC23,DBLP:journals/tois/GuoCBFCZC25,DBLP:conf/emnlp/SokliPKP25}, has drastically shifted the field.
MoE models~\cite{DBLP:journals/neco/JacobsJNH91} introduce a gating mechanism that dynamically routes inputs through specialized experts, inducing a structured partition of the embedding space whose boundaries reflect routing decisions rather than input features.
Therefore, explaining how expert routing shapes retrieval decisions requires moving beyond scalar feature importance towards interpretation of the embedding space geometry.
Research in neural ExIR has taken steps in this direction.
\citet{DBLP:conf/sigir/FernandoSA19} applied gradient-based attribution to interpret relevance predictions in neural rankers, \citet{DBLP:conf/ictir/VolskeBF0SHA21} provided an axiomatic perspective on neural ranking behavior, while \citet{DBLP:conf/ecir/WallatBAA23} investigated probing techniques for understanding learned representations in BERT-based rankers, presenting a representation-level analysis of neural IR models.
However, the geometry of the embedding spaces induced by these models remains underexplored.

Interpreting such a structure at the concept level also calls for methods that can associate embedding directions with human-interpretable semantic attributes.
CAVs~\cite{kim2018interpretabilityfeatureattributionquantitative,holistic_concepts} paired with sparse dictionary learning~\cite{bricken2023towards,gao2025scaling} provide faithful concept-level NLP explanations, making them ideal for MoE settings where experts induce subspace directions. 
Because this unsupervised approach requires post-hoc concept labeling~\cite{bricken2023towards,huben2024sparse}, manually scaling to thousands of concepts is labor-intensive. 
Consequently, automated LLM-based labeling, which generates concise labels from top-activating samples, has emerged as a promising, efficient alternative \cite{feldhus-kopf-2025-interpreting,huben2024sparse}.
Despite these works, their application to IR remains underexplored, with one notable exception employing supervised CAVs to improve personalized semantics in recommender systems~\cite{reccavs1}.

The two MoE-enhanced IR models most relevant to this work are SB-MoE \cite{DBLP:journals/corr/abs-2510-15683} and DenseC3 \cite{DBLP:conf/emnlp/SokliPKP25}.
Both models enhance a standard dense bi-encoder with a single MoE layer placed atop the encoder backbone, where six experts receive the query or document embedding produced by the backbone, and their outputs are fused into a single representation via a gating-weighted sum.
SB-MoE trains its router in an \textsc{unsupervised} manner, dynamically optimizing expert assignments based on the downstream IR tasks.
The router is a learned gating function that takes the query or document embedding as input and produces a weight vector over the experts, indicating the relevance of each expert's contribution to the given input.
Formally, let $x$ be the input embedding, $f_i(x)$ the output of the $i$-th expert, and $g_i(x)$ the weight assigned to the $i$-th expert by the gating function; the final refined representation $y$ is the weighted sum of all experts' outputs:
\begin{equation}
y = \sum_{i=1}^{n} f_i(x) \cdot g_i(x)
\end{equation}
During training, noisy Top-1 gating~\cite{DBLP:conf/iclr/ShazeerMMDLHD17} ensures that all experts are explored, while during inference, the weighted sum of all experts' outputs is used to produce the final representation.
DenseC3 instead trains its router with Cognitive Complexity labels derived from Bloom's Taxonomy~\cite{anderson2001taxonomy}, a six-level classification of texts based on their understandability and intended usage, providing a weakly \textsc{supervised} signal that directly informs expert assignment.
Rather than learning routing decisions from the retrieval signal alone, DenseC3's gating mechanism is implemented as a multi-label classifier pre-trained on external annotated datasets, producing complexity weights that route each input to the expert corresponding to each Cognitive Complexity level.

Data visualization techniques such as t-SNE~\cite{maaten2008visualizing} or UMAP~\cite{mcinnes2018umap} comprise popular tools for the visualization and inspection of embedding space structure. 
Because they are unsupervised, however, they can miss relevant structure in the projections. 
In contrast, DeepView~\cite{DBLP:conf/ijcai/0001HH20} utilizes a discriminative metric together with UMAP in order to emphasize user-relevant information. 
See Figure \ref{fig:sub1} and~\ref{fig:sub2} for an example in the IR setting, where regular UMAP does not reveal any structure, but DeepView (DiDi UMAP) does.
Formally, the discriminative distances used in DeepView emphasize changes in a discrete variable like a class label and are defined as shortest path integrals, which locally measure the divergence between probability distributions over the discrete variable on nearby positions. 
This is equivalent to the Fisher information metric~\cite{DBLP:conf/ijcai/0001HH20} and is approximated by 
\begin{equation} \label{eq:deepview}
    d(x,y) := \sum_{i=1}^{n}(1-\lambda)\,d_{JS}(f(p_{i-1}),f(p_i)) + \lambda\,d_S(p_{i-1},p_i)
\end{equation}
for two points $x,y$ in the original input space, $f$ the classifier outputting a probability distribution over classes, $d_{JS}$ the Jensen-Shannon metric, $d_S$ an unsupervised regularization metric, $p_i$ equidistant points on the straight line from $x$ to $y$, and $\lambda \in [0,1]$ balancing the discriminative and unsupervised components. 
This metric, together with UMAP, is referred to as discriminative dimensionality reduction (DiDi).

In this work, we propose ExpertLens, a post-hoc explainability framework for MoE-enhanced dense retrievers that leverages DeepView to project the embedding space through the lens of each MoE model's own routing behavior.
By deriving $f$ from the internal gating mechanism, the discriminative boundaries encoded in the DeepView projections correspond directly to expert routing decisions, yielding visualizations that reflect the subspace geometry induced by MoE in a way that generic dimensionality reduction methods cannot.

\section{The Proposed Framework}
This section motivates the adoption of DeepView as the visualization framework of ExpertLens, and the selection of SB-MoE and DenseC3 as a controlled contrastive pair of MoE-enhanced dense retrievers for investigating MoE-induced embedding space structure within the scope of post-hoc ExIR (Section~\ref{sec:motiv}).
We then describe the three phases of the ExpertLens framework (Section~\ref{sec:ExpertLens}).

\subsection{Motivation} \label{sec:motiv}
Dense retrievers project both queries and documents into a shared high-dimensional embedding space, where retrieval decisions are determined by geometric proximity.
When enhanced with MoE mechanisms, these models induce distinct expert-specific subspaces, effectively imposing a structured geometry that reflects the division of representational capacity across experts.
These additional subspaces, absent in standard dense retrievers, have important implications for retrieval, as the expert to which a document is routed influences the region of the embedding space it occupies and, consequently, the set of queries for which it can be retrieved.
Revealing the subspace geometry induced by MoE requires a visualization framework that encodes routing boundaries into its projection.
As shown in Figure~\ref{fig:disc_vs_reg}, conventional dimensionality reduction techniques can obscure the distinct expert subspaces, as they do not leverage the model's routing decisions.
DeepView addresses this directly, and within the ExpertLens setting, the classifier $f$ (Eq.~\ref{eq:deepview}) is derived from each model's internal gating mechanism, meaning the discriminative boundaries encoded in the projection correspond directly to expert routing decisions.
Investigating how the gating mechanism shapes embedding space structure requires a controlled setting in which architectural variation is minimized, and the routing strategy is the primary variable.
We therefore select SB-MoE~\cite{DBLP:journals/corr/abs-2510-15683} and DenseC3~\cite{DBLP:conf/emnlp/SokliPKP25} as a controlled contrastive pair.
Both models share the same single-layer MoE architecture with six experts and a gate-weighted summed output, differing solely in their gating mechanism (\textsc{unsupervised} in SB-MoE and \textsc{supervised} in DenseC3).
This architectural alignment ensures that any observed differences in embedding space structure or expert specialization are attributable to the gating mechanism rather than to architectural variation.

\begin{figure*}[t]
\centering

\begin{subfigure}[t]{0.32\textwidth}
    \centering
    \includegraphics[width=\linewidth]{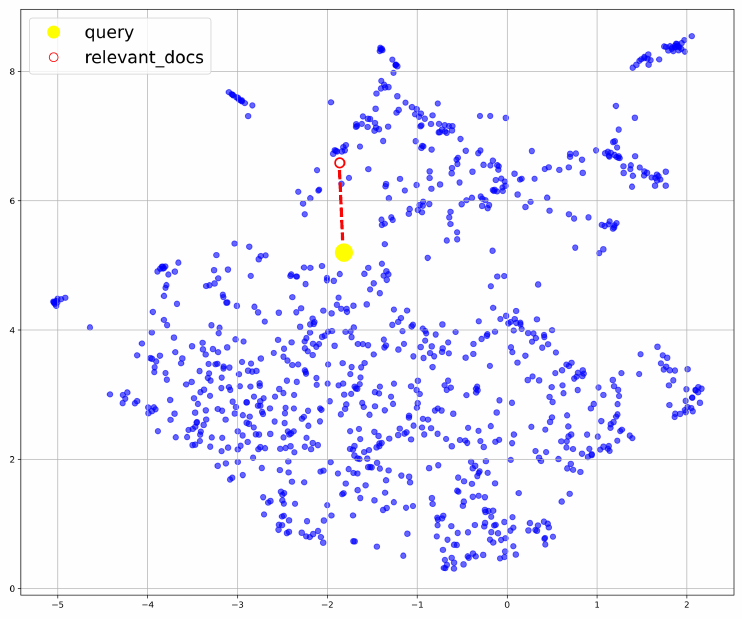}
    \caption{Fine-tuned (without MoE)}
    \label{fig:finetuned}
\end{subfigure}
\hfill
\begin{subfigure}[t]{0.32\textwidth}
    \centering
    \includegraphics[width=\linewidth]{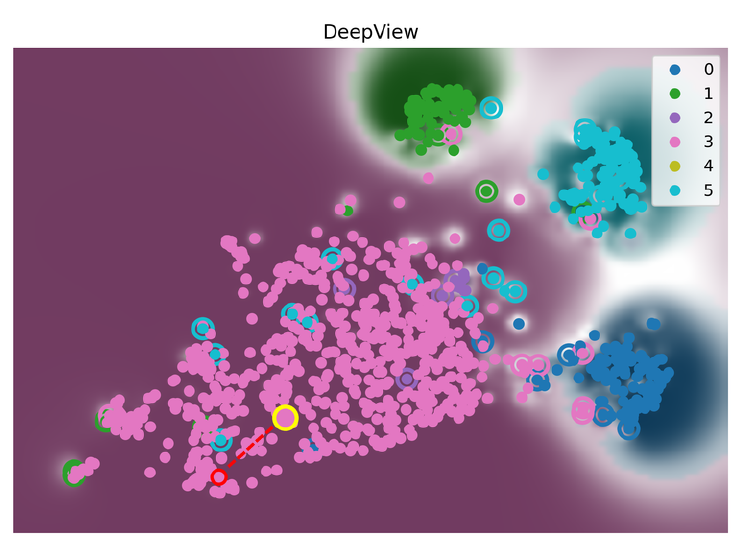}
    \caption{\textsc{Unsupervised} MoE}
    \label{fig:unsupervised}
\end{subfigure}
\hfill
\begin{subfigure}[t]{0.32\textwidth}
    \centering
    \includegraphics[width=\linewidth]{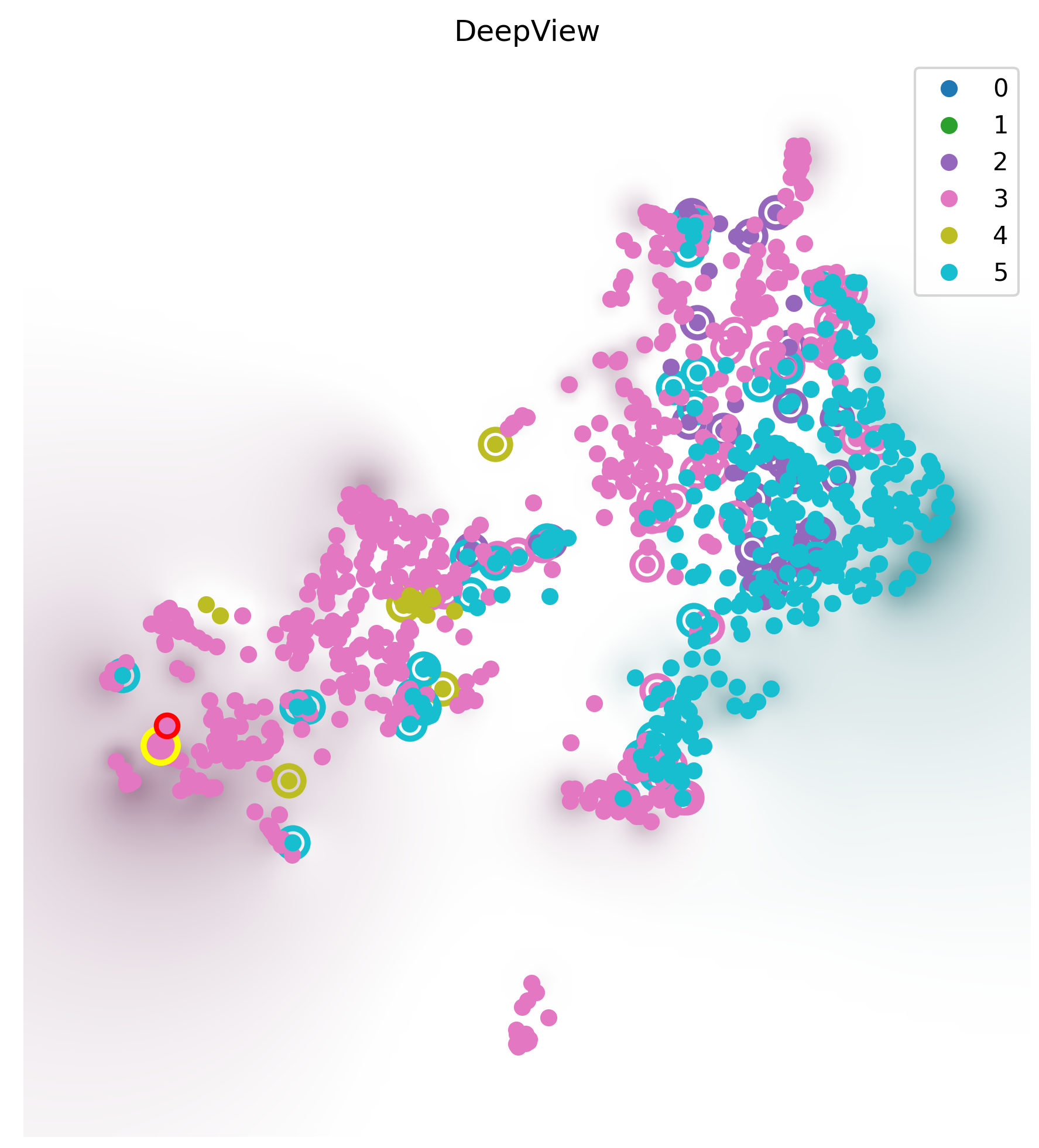}
    \caption{\textsc{Supervised} MoE}
    \label{fig:supervised}
\end{subfigure}

\caption{UMAP (a) and DeepView (b \& c) visualizations of the embedding space for the same query (yellow circle) and its relevant document (red circle) across three model variants: (a) Fine-tuned (without MoE), (b) \textsc{Unsupervised} MoE (SB-MoE), and (c) \textsc{Supervised} MoE (DenseC3). Each plot shows the top-1000 ranked documents retrieved for this query by each model, with colors indicating expert assignment in MoE variants.}
\label{fig:alignment}
\end{figure*}
\subsection{ExpertLens}\label{sec:ExpertLens}
Figure~\ref{fig:framework} illustrates the ExpertLens framework, which consists of three main phases:
(I) \textit{MoE Induced Embedding Space Structure and Analysis}, where discriminative projections of the embedding space are computed, and its geometric structure is analyzed;
(II) \textit{Subspace Disentanglement and Expert Specialization}, 
where the expert-induced subspaces emerging from the projected embedding space are characterized and associated with expert specialization; and
(III) \textit{Concept Extraction and Expert Attribution}, 
where CAVs are automatically extracted and mapped to individual experts, linking their specialization to interpretable semantic concepts.

\subsubsection{Phase I} \label{sec:phase1}
In MoE-enhanced dense retrievers, a learned gating mechanism assigns each input representation to one or more specialized experts, effectively partitioning the embedding space according to the model's routing behavior. 
DeepView projects data points using the model's decision boundaries, combining feature-wise similarity with classifier-induced distance to emphasize directions in the embedding space where model predictions change~\cite{DBLP:journals/corr/abs-2403-18872}. 
The classifier distilled for DeepView\footnote{Technical details of the classifier and its distillation procedure are reported in Section~\ref{sec:phase1setup}.} ($f$ in Eq.~\ref{eq:deepview}) is derived from each MoE model's internal gating mechanism, ensuring that the discriminative projection reflects expert assignment boundaries rather than externally imposed class labels.
From the derived DeepView visualizations, we observe that MoE-enhanced variants progressively reposition relevant documents into closer relative proximity to their corresponding queries compared to standard dense retrievers, suggesting that expert routing induces a more structured query neighborhood within the embedding space (Figure~\ref{fig:alignment}).
To quantify how expert routing affects the organization of the embedding space, we analyze the neighborhood structure around each query from two complementary perspectives.
First, we measure the proportion of relevant documents appearing among the $k$ nearest neighbors of each query $q$ by dot product, equivalent to Precision@$k$:
\begin{figure}[t]
    \centering
    \includegraphics[width=0.7\columnwidth]{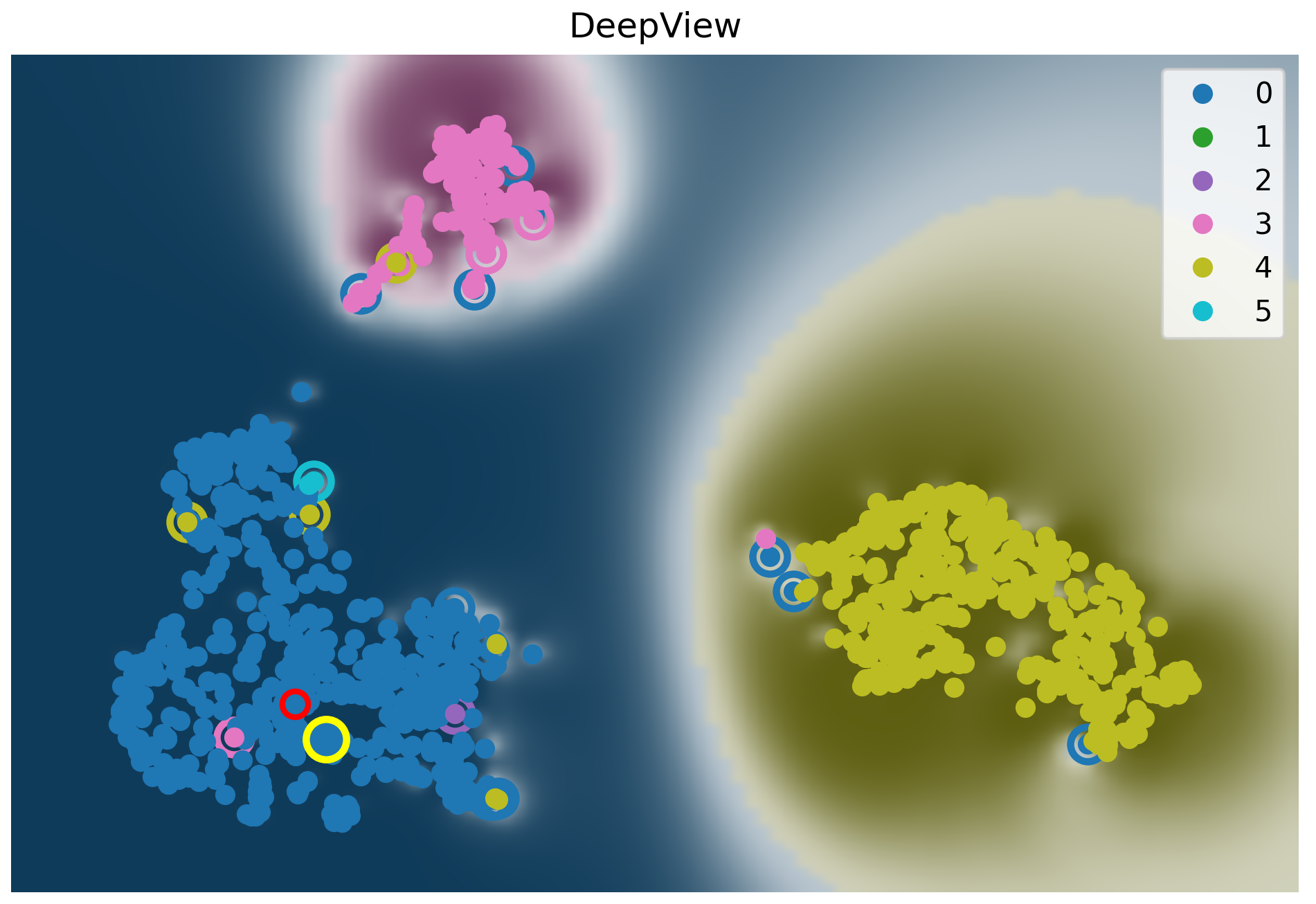}
    \vspace{0.5em}
    \includegraphics[width=0.72\columnwidth]{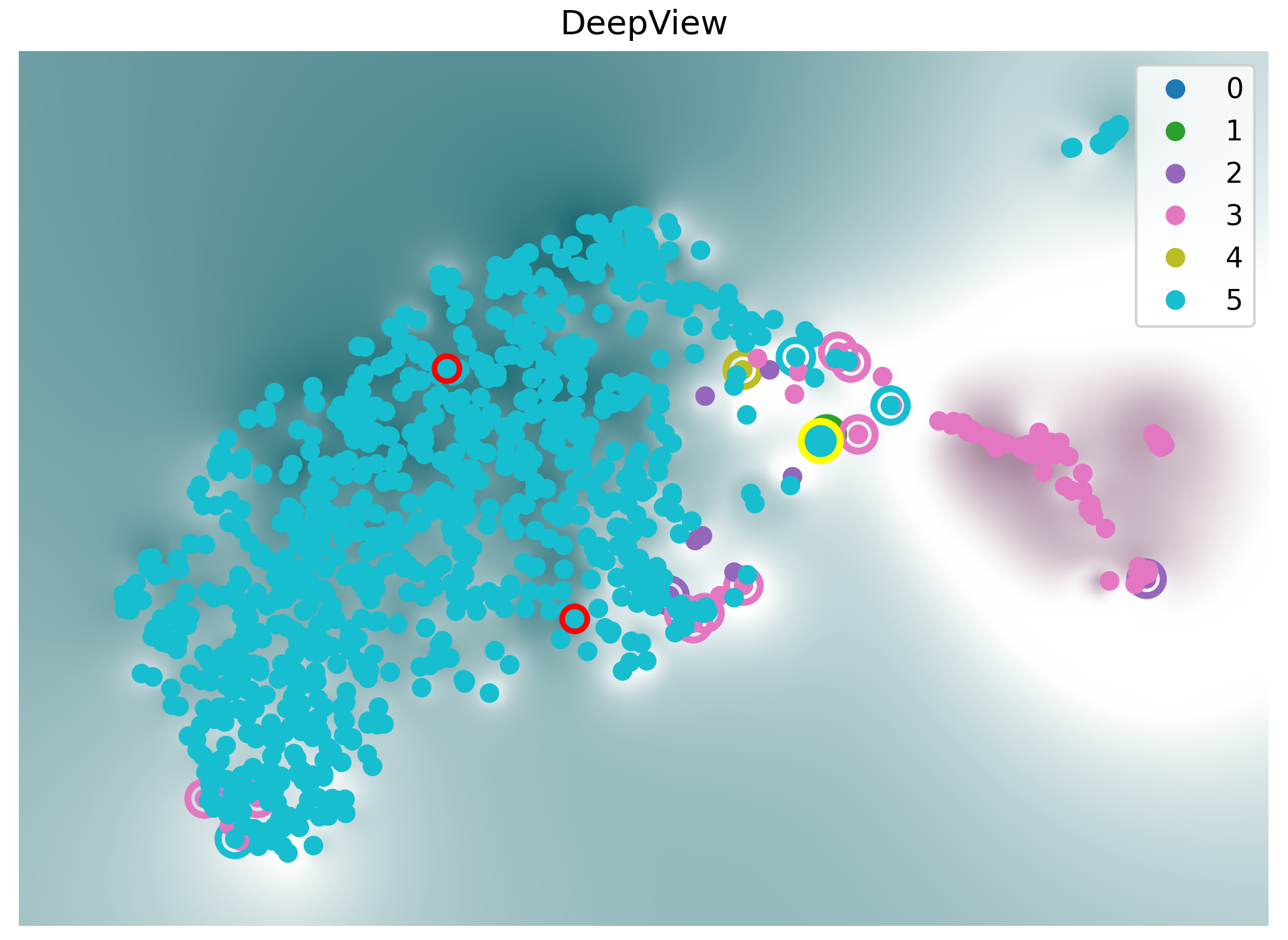}
    \caption{DeepView plot of Query ID: ``181476" from MSMARCO by the \textsc{Unsupervised} MoE variant (top), and Query ID: ``5a7199725542994082a3e88f" from HotpotQA by the \textsc{Supervised} MoE variant (bottom).}
    \label{fig:example_exp2}
\end{figure}
\begin{equation}
\text{P@}k(q) = \frac{|D^+ \cap N_k(q)|}{|N_k(q)|}
\end{equation}
where $N_k(q)$ denotes the set of $k$ nearest neighbors (i.e., documents) of $q$ and $D^+$ its set of relevant documents.
This reflects the relative positioning of relevant documents within the immediate query neighborhood.
Second, we measure the absolute proximity of each relevant document $d^+$ to its corresponding query $q$ via cosine distance:
\begin{equation}
\text{$cos_{dist}$}(q, d^+) = 1 - \frac{q \cdot d^+}{\|q\| \cdot \|d^+\|}
\end{equation}
Together, these two measures allow us to assess whether MoE induce a better-structured embedding space by positioning relevant documents into better-defined geometric neighborhoods with respect to their queries, comparing the \textsc{unsupervised} and \textsc{supervised} MoE implementations with standard \textsc{fine-tuned} retrievers.

\subsubsection{Phase II}\label{sec:phase2}
Based on the DeepView projections obtained in Phase I, we conduct a linguistic and statistical analysis of each expert-associated embedding subspace to characterize the textual properties each expert specializes in.
For the linguistic profiling to yield meaningful expert-specific outcomes, the top-$k$ retrieved documents of a given query must form clearly separated, expert-aligned clusters in the embedding space.
Queries whose retrieved documents are spread uniformly across experts or occupy overlapping regions provide no discriminative basis for subspace characterization.
Rather than manually selecting representative queries or analyzing all queries indiscriminately, we therefore introduce an automatic query scoring procedure that identifies queries whose top-$k$ retrieved documents exhibit well-structured, expert-aligned clustering in the embedding space (Figure~\ref{fig:example_exp2}).
For each query $q$, we obtain the two-dimensional DeepView projections of its top-1000 retrieved documents from Phase I, together with the expert ID each document was routed to.
We then apply DBSCAN~\cite{ester1996density} to the 2D projected space using Euclidean distance, discovering natural density-based clusters in the discriminative projection.
DBSCAN makes no assumption about cluster shape or count, grouping documents that lie within $\varepsilon$ Euclidean distance of at least $\textit{min\_samples}$ neighbors, and labeling documents that cannot be attached to any dense region as noise.
We then measure how well the discovered clusters align with the expert assignments using the Adjusted Rand Index (ARI)~\cite{hubert1985comparing}, computed between the DBSCAN cluster labels and the expert IDs, restricted to non-noise documents:
\begin{equation}
\text{ARI} = \frac{\text{RI} - \mathbb{E}[\text{RI}]}{\max(\text{RI}) - 
\mathbb{E}[\text{RI}]} \in [-1, 1]
\end{equation}
where $\text{RI}$ is the Rand Index between the two partitions, and $\mathbb{E}[\text{RI}]$ is its expected value under random assignments.
An ARI of $+1$ indicates that DBSCAN clusters perfectly mirror expert assignments, $0$ indicates no better than random agreement, and $-1$ indicates systematic anti-correlation.
To account for the reliability of the clustering itself, we weight the ARI by the \textit{core ratio} $\rho$, defined as the fraction of non-noise documents:
\begin{equation}
\rho = \frac{|\{d : \text{label}(d) \neq -1\}|}{|D_q|}
\end{equation}
where $D_q$ is the set of top-1000 retrieved documents for query $q$.
A high $\rho$ indicates that most retrieved documents form compact, high-density clusters, making the ARI estimate more trustworthy.
The composite clustering quality score for query $q$ is then:
\begin{equation} \label{cluster_score}
\text{score}(q) = \rho \cdot \max(0, \text{ARI})
\end{equation}
This score is high only when the retrieved documents form compact clusters within the embedding space, and the expert routing correctly mirrors those clusters.
Queries whose top-$k$ documents are spread uniformly in embedding space, or whose expert labels disagree with the natural density structure, score near $0$.
Queries for which DBSCAN finds fewer than two clusters are assigned a score of $-1$ and excluded from the analysis.
We compute this score across all queries in each dataset and retain the top-100 queries per dataset per model for the subsequent linguistic profiling, providing a substantial and automatically grounded empirical basis for the qualitative expert specialization analysis.
Finally, for each selected query, we profile the subspace associated with each active expert using five corpus-level linguistic metrics that are lightweight, domain-agnostic, and comparable across corpora of varying sizes~\cite{DBLP:conf/lrec/BankRS12,DBLP:conf/sigir/LiuJYWCCG23,DBLP:conf/sac/PallucchiniZ0C25,senter1967automated}.
The \textit{Automated Readability Index} ($ari$) estimates text complexity from average word and sentence length.
\textit{Shannon's Entropy} ($H$) measures the average information content of tokens, normalized by vocabulary size, capturing the uniformity of the token distribution.
The \textit{Relative Vocabulary Size} ($R_{\text{Voc}}$) quantifies lexical diversity as the ratio of vocabulary size to the number of non-function tokens.\footnote{The function words $N_f$ defined in \citet{DBLP:conf/lrec/BankRS12}: the, a, an, he, him, she, her, they, us, we, them, it, his, to, on, above, below, before, from, in, for, after, of, with, at, and, or, but, nor, yet, so, either, neither, both, whether.}
\textit{Corpus Predictability} ($CP$) captures sequential token dependencies via a first-order Markov model, where higher values indicate more predictable token sequences.
Finally, \textit{Average Sentence Length} ($L_S$) measures the mean number of tokens per sentence.
Expert specialization is assessed per subspace and contrasted against the top-1000 retrieval pool and the full corpus baseline.

\subsubsection{Phase III} \label{sec:phase3}
While Phase II characterizes expert subspaces through linguistic properties, Phase III identifies the semantic concepts that structure the embedding space and evaluates how consistently these concepts align with expert routing decisions.
To this end, we extract CAVs via Sparse Autoencoder (SAE) decomposition~\cite{holistic_concepts}, and subsequently label each concept using an LLM \cite{feldhus-kopf-2025-interpreting}.
We decompose the $n$-document embedding matrix $A \in \mathbb{R}^{n \times d}$ into a sparse weight matrix $U \in \mathbb{R}^{n \times c}$ and a concept dictionary $V \in \mathbb{R}^{d \times c}$, where $c \gg d$ forms an overcomplete basis:
\begin{equation}
(U, V) = \arg\min_{U,V} \|A - UV^\top\|_F^2
\end{equation}
Each column of $V$ defines a learned concept direction in the embedding space, and the non-zero entries of each row of $U$ identify which concepts are active for a given document.
The overcomplete basis encourages sparsity, ensuring that each document is described by a small, interpretable subset of concepts.
To assign human-interpretable labels to learned concepts, we embed the queries to obtain $A_q \in \mathbb{R}^{n_q \times d}$ and project them onto the concept dictionary to obtain query activations $U_q \in \mathbb{R}^{n_q \times c}$.
For each concept $c_i$, we rank queries by their activation score and pass the top-20 to an LLM requesting a common semantic theme.
Queries are preferred over documents for labeling as they are shorter and topically focused, yielding cleaner concept descriptions.
Long documents often span multiple topics, risking the LLM latching onto incidental rather than defining semantic properties of a concept.
We acknowledge that this simplification may cause the LLM to capture associations that are cleaner but potentially narrower than what the full document text would reveal; we therefore treat concept labels as indicative semantic summaries rather than exhaustive characterizations.
The expert-concept association is determined by aggregating the non-zero concept activations per expert across the full document corpus, subsequently visualized as a Sankey diagram to reveal how semantic concepts propagate across experts (Figure~\ref{fig:sankey_concepts}).
To evaluate whether the embedding space organizes itself around semantically coherent structure, we define the $Q_{concept}(n)$ error to measure how consistently the top-$n$ active concepts of a document are shared with its nearest neighbors.
A low error indicates that geometrically proximate documents activate similar concepts, confirming that embedding subspaces are structured along semantically coherent directions rather than organized arbitrarily.
Conversely, a high error suggests that neighboring documents do not share dominant concepts, implying that the subspace geometry is not driven by semantic content.
For a document $d$, let $S_d^{(n)}$ denote its top-$n$ concept activations and $N_K(d)$ its $K$ nearest neighbors by cosine similarity:
\begin{equation}
Q_{concept}(n) = \frac{1}{|Q| \cdot K} \sum_{d \in Q} \sum_{j \in N_K(d)} 
\left(1 - \frac{|S_d^{(n)} \cap S_j^{(n)}|}{n}\right)
\end{equation}
where $Q$ is a set of sampled documents.
We evaluate $Q_{concept}$ across supervised and unsupervised MoE variants to assess how routing affects the semantic coherence of the induced embedding space.

\section{Experimental Setup} \label{sec:experiments}
In this section, we report the empirical evaluation conducted to address the following research questions (RQs):
\textbf{RQ1.} How can MoE enhancements in dense retrievers affect the structure and immediate query neighborhood within the embedding space (Phase I)?
\textbf{RQ2.} What textual characteristics of their associated subspaces define expert specialization (Phase II)? 
\textbf{RQ3.} Can CAVs reveal interpretable semantic concepts associated with the specialization of individual experts (Phase III)?

\subsection{Datasets, Models \& Baselines}
We conduct experiments on five publicly available benchmarks (Table~\ref{tab:datasets}) spanning two IR tasks: (i) passage retrieval and (ii) open-domain question answering (Q\&A) formulated as a search task~\cite{DBLP:conf/acl/ChenFWB17}.
TREC DL 19 \& 20 share the MSMARCO corpus and provide dense relevance judgments, offering a rich signal for the embedding space structure analysis in Phase I.
MSMARCO, NQ, and HotpotQA (all part of BEIR~\cite{DBLP:journals/corr/abs-2104-08663}) are used across all three phases of ExpertLens, providing corpus scale and topical diversity required for the per-expert subspace analysis in Phases II and III.
We evaluate ExpertLens on two MoE-enhanced dense retrievers, SB-MoE~\cite{DBLP:journals/corr/abs-2510-15683} and DenseC3~\cite{DBLP:conf/emnlp/SokliPKP25}, instantiated on four encoder backbones of varying scale and pre-training strategy (Table~\ref{tab:backbones}).
TinyBERT and BERT-Small are obtained via knowledge distillation~\cite{DBLP:journals/corr/RomeroBKCGB14}.
BERT-Base uses standard masked language modeling pre-training, and ColBERT is further fine-tuned with a late-interaction retrieval objective, making it the only backbone with IR-specific pre-training.
For each backbone, we compare three model variants: (i) \textsc{Fine-Tuned} (without MoE), (ii) \textsc{Unsupervised} (SB-MoE), and (iii) \textsc{Supervised} (DenseC3).
The \textsc{Fine-Tuned} baseline is a standard dense bi-encoder trained without any MoE layer, using the same training data, number of epochs, learning rate, and batch size as the MoE variants, following the configurations reported in the original works~\cite{DBLP:journals/corr/abs-2510-15683,DBLP:conf/emnlp/SokliPKP25}, ensuring that observed differences in embedding space structure are attributable to MoE routing rather than training differences.
Phase I is conducted across all four backbones.
Phases II and III, which require per-query analysis and SAE training, are conducted on TinyBERT to maintain comparability while managing computational costs.

\begin{table}[t]
\caption{Dataset statistics. The average number of relevance judgments per query is indicated in parentheses.}
\label{tab:datasets}
\centering
\resizebox{1\linewidth}{!}{%
\begin{tabular}{l|c|c|c}
\hline
\textbf{Dataset} & \textbf{Corpus Size} & \textbf{Train Queries} & \textbf{Test Queries} \\
\hline
MSMARCO v1 passage corpus (MSMARCO~\cite{DBLP:conf/nips/NguyenRSGTMD16}) & 8.8M passages & 532k (1.1 avg rel.) & 7k \\
TREC Deep Learning Track 2019 (TREC DL 19~\cite{DBLP:journals/corr/abs-2003-07820}) & 8.8M passages & 503k (215 avg rel.) & 43 \\
TREC Deep Learning Track 2020 (TREC DL 20~\cite{DBLP:journals/corr/abs-2102-07662}) & 8.8M passages & 503k (211 avg rel.) & 54 \\
Natural Questions (NQ~\cite{kwiatkowski2019NQ}) & 2.6M passages & 132k (1.2 avg rel.) & 3.5k \\
HotpotQA~\cite{yang-etal-2018-hotpotqa} & 5.2M documents & 85k (2 avg rel.) & 7.4k \\
\hline
\end{tabular}
}
\end{table}
\begin{table}[t]
\caption{Backbone bi-encoder models statistics. Parameter count, embedding size, and pre-training strategy.}
\label{tab:backbones}
\centering
\resizebox{1\linewidth}{!}{%
\begin{tabular}{l|c|c|c}
\hline
\textbf{Model} & \textbf{Parameters} & \textbf{Embedding Size} & \textbf{Pre-Training Strategy} \\ 
\hline
TinyBERT~\cite{jiao-etal-2020-tinybert} & 14.5M & 312 & Knowledge Distillation \\
BERT-Small~\cite{turc2019well} & 29M & 512 & Knowledge Distillation \\
BERT-Base~\cite{DBLP:conf/naacl/DevlinCLT19} & 110M & 768 & Self-Supervised Learning \\
ColBERT~\cite{DBLP:conf/sigir/KhattabZ20}   & 110M & 768 & IR Fine-Tuned \\
\hline
\end{tabular}
}
\end{table}
\begin{figure}[t]
    \centering
    \begin{subfigure}{0.32\columnwidth}
        \centering
        \includegraphics[width=\linewidth, trim=0cm 5cm 0cm 1.5cm, clip]{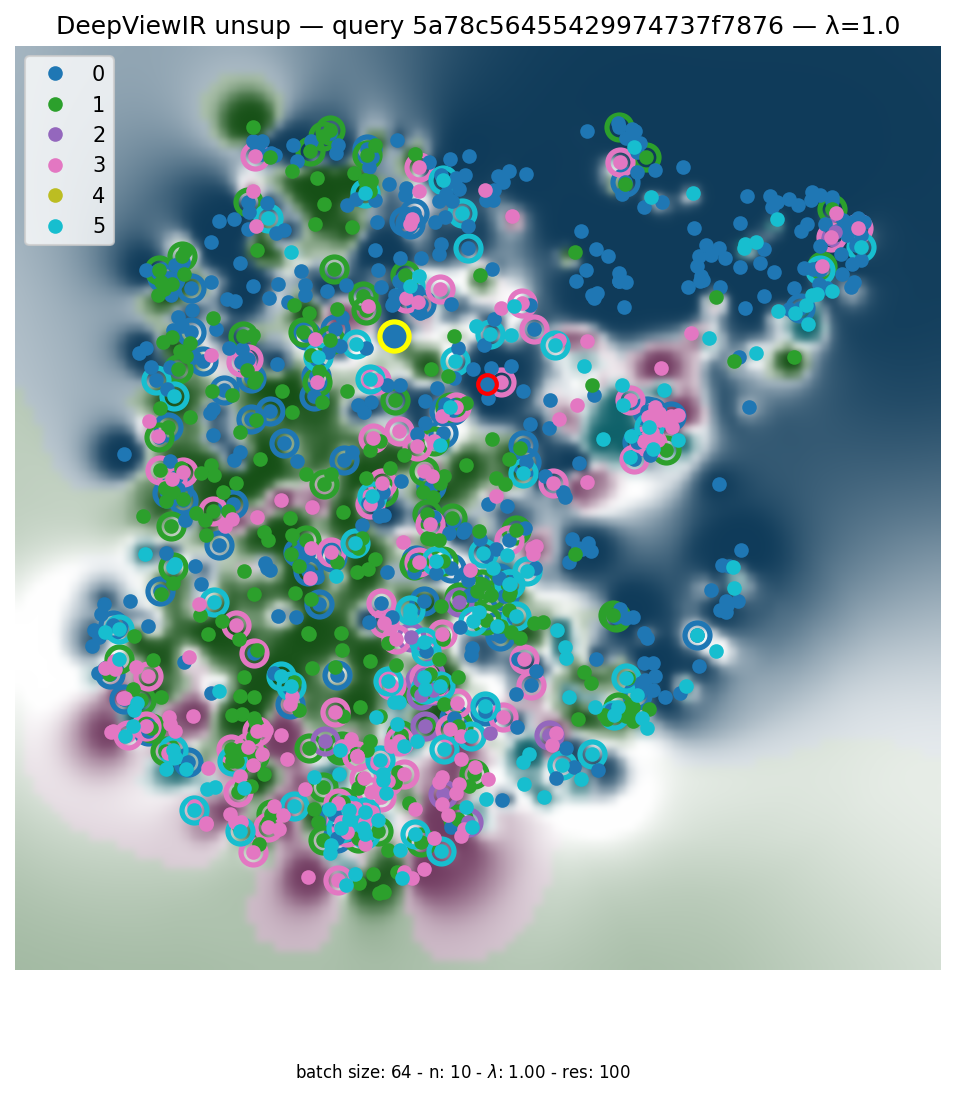}
        \caption{Regular UMAP}
        \label{fig:sub1}
    \end{subfigure}
    \hfill
    \begin{subfigure}{0.32\columnwidth}
        \centering
        \includegraphics[width=\linewidth, trim=0cm 5cm 0cm 1.5cm, clip]{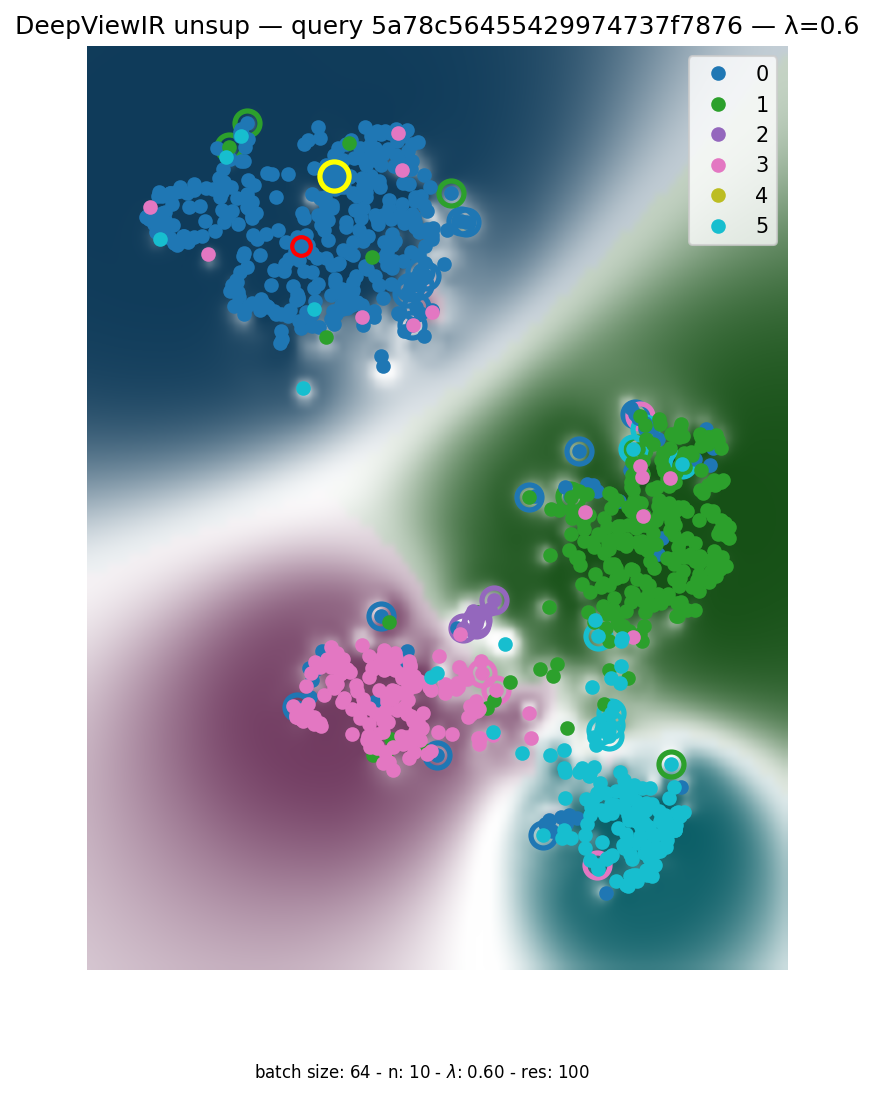}
        \caption{DiDi UMAP}
        \label{fig:sub2}
    \end{subfigure}
    \hfill
    \begin{subfigure}{0.32\columnwidth}
        \centering
        \resizebox{\linewidth}{!}{%
        \begin{tabular}{cccc}
            \toprule
            $\lambda$ & $Q_\text{data} \uparrow$ & $Q_{\neg data} \uparrow$ & $Q_\text{knn} \downarrow$ \\
            \midrule
            0.0 & 95.6\% & 94.7\% & 0.0929 \\
            0.1 & 96.3\% & 98.0\% & 0.0979 \\
            0.2 & 94.6\% & 95.7\% & 0.0989 \\
            0.3 & 95.9\% & 96.7\% & 0.0939 \\
            0.4 & 96.4\% & 98.0\% & 0.0899 \\
            0.5 & 96.6\% & 95.3\% & 0.0959 \\
            0.6 & 97.4\% & 95.3\% & 0.0959 \\
            0.7 & 96.7\% & 96.0\% & 0.1009 \\
            0.8 & 94.7\% & 89.0\% & 0.1259 \\
            0.9 & 87.7\% & 67.8\% & 0.2028 \\
            1.0 & 80.1\% & 48.8\% & 0.4545 \\
            \bottomrule
        \end{tabular}
        }
        \caption{$\lambda$ values span}
        \label{tab:lambda_metrics}
    \end{subfigure}
    \caption{Discriminative UMAP (center) reveals clear clusters of documents processed by the same experts for a given query compared to regular UMAP (left). Evaluation metrics across $\lambda$ values for query \texttt{5a78c56455429974737f7876} (right). \textsc{Unsupervised} MoE on HotpotQA.}
    \label{fig:disc_vs_reg}
\end{figure}
\input{phase1}
\subsection{Phase I Setup} \label{sec:phase1setup}
The DeepView visualizations are produced by distilling a classifier from the gating mechanism of each model.
The classifier consists of two down-projection layers, the first reducing the input from the backbone embedding size to half, and the second reducing it further to the number of experts (six).
It is trained with Kullback–Leibler divergence loss against the expert logits for 30 epochs on the training set of each dataset, with a 5\% validation split, retaining the checkpoint with the lowest validation loss.
To select the optimal $\lambda$ balancing discriminative and unsupervised components in DeepView (Eq.~1), we perform a grid search over $\lambda \in [0, 1]$ with a step of 0.1.
For each value, we evaluate the resulting projection using three metrics~\cite{DBLP:conf/ijcai/0001HH20}: $Q_{data}$ and $Q_{\neg data}$, which assess how well the scatter plot aligns with the classification function for the plotted and background points respectively, and $Q_{knn}$, which measures the extent to which the five nearest neighbors of a given point share the same expert.
Figure~\ref{tab:lambda_metrics} reports the full grid search results for a representative query.
High values ($\lambda \geq 0.8$) increasingly suppress the discriminative signal, with $\lambda = 1.0$ reducing to regular UMAP, which fails to reflect expert assignment boundaries as evidenced by $Q_{knn}$ deteriorating sharply to 0.455 (Figure~\ref{fig:sub1}).
At $\lambda = 0.6$, $Q_{data}$ maximizes (97.4\%) while $Q_{\neg data}$ (95.3\%) and $Q_{knn}$ (0.096) remain stable, and is therefore applied uniformly across all models, backbones, and datasets (Figure~\ref{fig:sub2}).
The choice of Precision@$k$ cutoffs is grounded in the relevance judgment density of each dataset.
TREC DL~19 \& 20 provide an average of over 200 relevance judgments per query; we therefore evaluate at $k \in \{20, 100\}$, corresponding to approximately 10\% and 50\% of their average relevance count, respectively, to capture meaningful portions of the relevant document pool within the query neighborhood.
MSMARCO, NQ, and HotpotQA provide sparse judgments, with an average of only 1-2 relevant documents per query; hence, we select $k \in \{1, 2\}$ accordingly.
We report Precision@$k$ to capture improvements in the \textit{relative} query neighborhood structure by measuring whether relevant documents are positioned closer to the query compared to the remaining retrieved documents.
Cosine distance assesses the \textit{absolute} geometric proximity between queries and their relevant documents, and reveals whether the overall geometric spread of the embedding space contracts or expands under expert routing.
Together, these two metrics allow us to characterize how MoE reshape the embedding space.
We opt for this formulation because our goal is to evaluate the quality of the induced embedding spaces, rather than mere retrieval performance of the examined MoE models, which is already reported in the original works~\cite{DBLP:journals/corr/abs-2510-15683,DBLP:conf/emnlp/SokliPKP25}.

\subsection{Phase II Setup}
For each query, we apply DBSCAN~\cite{ester1996density} to the two-dimensional DeepView projections of the top-1000 retrieved documents obtained in Phase I, using Euclidean distance as the metric.
The maximum Euclidean distance between two documents for one to be considered in the neighborhood of the other is defined by $\varepsilon$, and the minimum number of documents required to form a dense region (core point) is defined by $\textit{min\_samples}$.
We set $\varepsilon = 0.25$ and $\textit{min\_samples} = 7$, selected via a small grid search to ensure that at least 100 queries with well-structured, expert-aligned partitions are retained per dataset per MoE variant for the subsequent linguistic profiling.
The composite clustering quality score (Eq.~\ref{cluster_score}) is computed for every query in each setting.
Queries for which DBSCAN finds fewer than two clusters are assigned a score of $-1.0$, while all remaining queries are scored by $\rho \cdot \max(0, \text{ARI})$ and ranked accordingly.
The ARI between DBSCAN cluster labels and expert IDs is computed exclusively on non-noise documents (i.e., documents not labeled $-1$ by DBSCAN), so that they do not distort the cluster-to-expert alignment signal.
The top-100 queries scored per dataset per MoE variant are then retained for linguistic profiling.
For each retained query, the textual characteristics are computed separately for the set of documents assigned to each active expert, treating it as an independent corpus, and mean values are reported across the 100 queries.
Expert specialization is assessed per subspace and contrasted against the top-1000 retrieval pool and the full corpus baseline.
We consider as active those experts handling at least 10 documents of the top-1000 retrieved for a given query on average across the analyzed queries.

\subsection{Phase III Setup}
We train a Top-$k$ SAE~\cite{gao2025scaling} with a concept dictionary of twice the embedding size (624 atoms for TinyBERT's 312-dimensional embeddings) and a 90\% sparsity penalty, on document embeddings from both MoE variants with a learning rate of $1e-3$ and batch size of 64 for 50 epochs. 
We compared its $R^2$ performance with an Archetypal Top-$k$ SAE with 10000 archetypes and the same hyperparameters; ultimately, we selected a vanilla Top-$k$ due to its better performance \cite{fel2025archetypalsaeadaptivestable}.
Concepts are labeled using Gemma-3-27b~\cite{gemmateam2025gemma3technicalreport} by providing the top-20 most activating queries per concept with a zero-shot prompt requesting a common semantic theme (prompt details and discovered example concepts in Appendix~\ref{app:prompt}).
We evaluate $Q_{concept}$ at $n$ corresponding to 1\%--10\% of the dictionary size (6--62 concepts), spanning from the single most dominant concept to the full top-$k$ activation set, with $K = 10$ nearest neighbors.
For 40 runs, we sample 20000 points and compute the $Q_{concept}$ for each point and report the average. 
The expert-concept association is determined by the non-zero concept activations per expert aggregated across the full document corpus.

\section{Results \& Discussion}
This section presents the results of the empirical evaluations conducted to address our Research Questions.

\input{phase2_unsup}
\input{phase2_sup}
\subsection{RQ1: Embedding Space Structure}
Figure~\ref{fig:alignment} presents DeepView visualizations of the embedding space derived from the three examined models for the same query and its relevant document, along with the top-1000 retrieved documents by each model.
A visual inspection reveals a progressive improvement in embedding space structure across model variants.
The relevant document (red circle) is geometrically distant from the query (yellow circle) in the \textsc{Fine-Tuned} embedding space, whereas MoE-enhanced variants reposition it closer to the query, suggesting that expert routing induces a more structured query neighborhood in which semantically related content is drawn into tighter relative proximity.
Table~\ref{tab:results_knn} reports mean Precision@$k$ evaluated for all three models with four different backbone bi-encoders across five IR benchmarks at various cutoffs (see Section~\ref{sec:phase1setup} for more details).
The results corroborate the qualitative observations from the DeepView visualizations across all backbones, i.e., the MoE-enhanced variants consistently achieve higher P@$k$ than \textsc{Fine-Tuned}.
This finding indicates that expert routing reduces the \textit{relative} distance between queries and their relevant documents compared to the remaining top-1000 retrieved pool in the embedding space.
The \textsc{Supervised} variant generally yields stronger results, with a marked 14.78\% gain over \textsc{Fine-Tuned} on NQ with TinyBERT at $k=1$.
Gains are most pronounced at tighter cutoffs (P@1), indicating that expert routing draws relevant documents into the immediate geometric neighborhood of the query rather than merely improving the broader embedding space structure, for a higher proportion of queries.
These findings are further examined through the mean cosine distance ($\text{cos}_{dist}$) between queries and their relevant documents.
Interestingly, MoE-enhanced variants consistently yield higher cosine distances than \textsc{Fine-Tuned}, suggesting that expert routing increases the overall geometric spread of the embedding space rather than compressing it, despite the reported improvements in retrieval effectiveness.
Notably, the increase in cosine distance is most pronounced for the \textsc{Supervised} variant, which simultaneously achieves the strongest P@$k$ gains, confirming that improved \textit{relative} neighborhood structure and increased \textit{absolute} spread co-occur as a direct consequence of expert-induced subspace partitioning.
Taken together, these results indicate that MoE routing consistently improves the relative positioning of relevant documents within the immediate query neighborhood as well as the overall embedding space structure for retrieval, while inducing a broader geometric spread across domains, IR tasks, and bi-encoders of varying scale and pre-training paradigms.

\subsection{RQ2: Expert Specialization}
\begin{figure}[t]
    \centering
    \begin{subfigure}[t]{\columnwidth}
        \centering
        \includegraphics[width=0.49\linewidth]{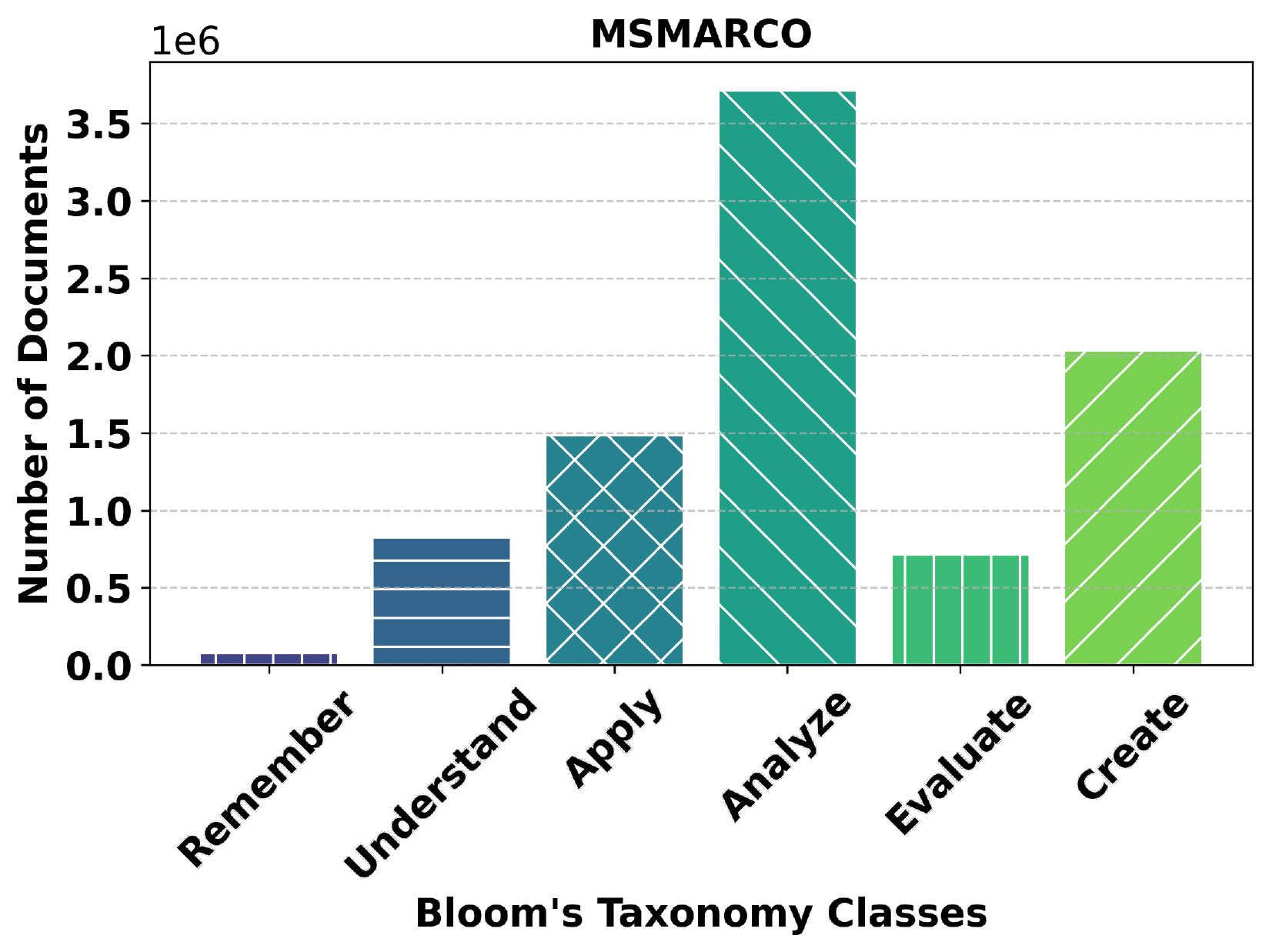}
    \end{subfigure}
    

    \begin{subfigure}[t]{\columnwidth}
        \centering
        \includegraphics[width=\linewidth]{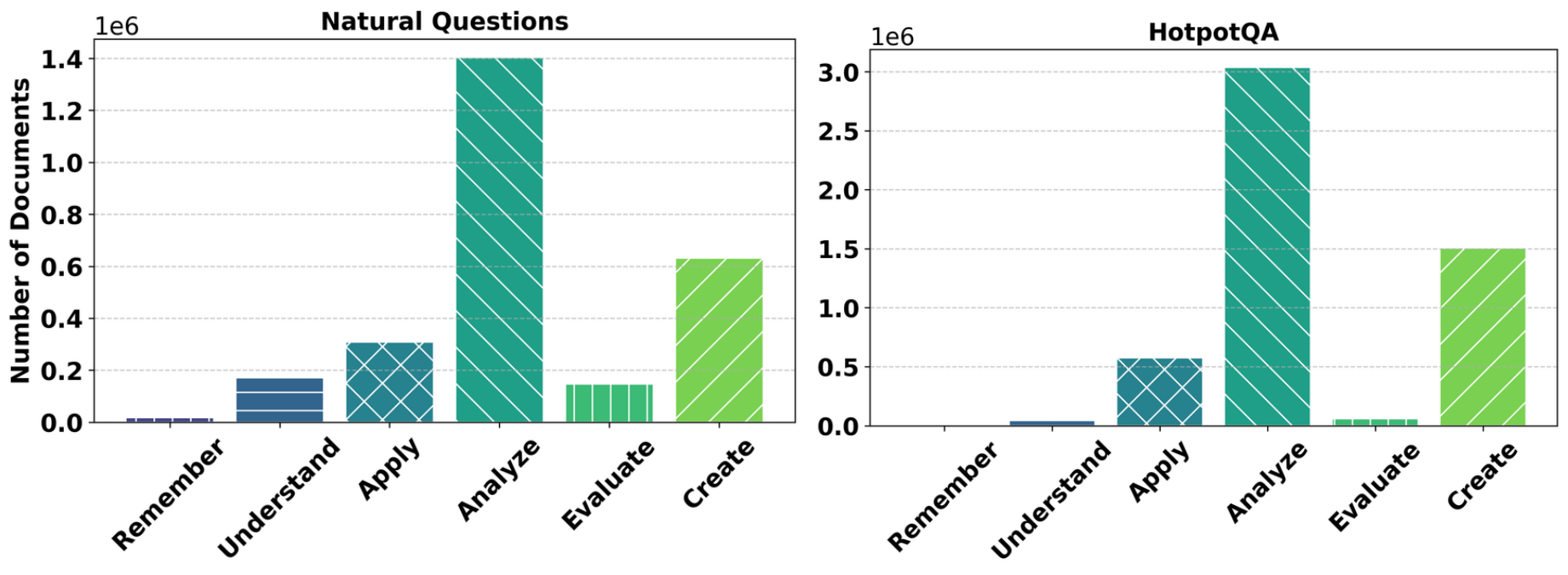}
    \end{subfigure}
    \caption{Expert assignment distributions on all datasets as reported in the original paper of the \textsc{supervised} MoE variant (DenseC3)~\cite{DBLP:conf/emnlp/SokliPKP25}.}
    \label{fig:distributions}
\end{figure}

Tables~\ref{tab:phase2_unsup} \&~\ref{tab:phase2_sup} report the linguistic profiles of representative expert subspaces for both MoE variants, illustrating the expert specialization patterns.
Across all three datasets, the \textsc{Unsupervised} variant exhibits consistent and interpretable expert specialization in terms of textual characteristics.
For every analyzed query, there is a \textit{dominant} expert that absorbs the majority of the top-1000 retrieved documents and scores lowest on lexical diversity ($R_{Voc}$), Shannon's entropy ($H$), and corpus predictability ($CP$), remaining close to the top-1000 and full corpus baselines.
In contrast, \textit{minority} experts consistently achieve the highest values on these three metrics, capturing lexically rich subspaces.
This inverse relationship between routing load and lexical richness holds without exception for all analyzed queries across datasets and IR tasks, suggesting that the unsupervised gating mechanism induces high-volume experts that produce general-purpose subspaces for structurally average documents, while low-volume experts capture text with more complex textual characteristics.
The expert specialization boundary is therefore formed by the intrinsic linguistic properties of the documents rather than by their semantic content.
The \textsc{Supervised} variant mirrors the specialization tendencies observed in the \textsc{Unsupervised} case, with one key difference.
The additional weak supervision signal produces a more balanced load distribution among experts, consistent with the routing distributions reported in the original work (Figure~\ref{fig:distributions}), and indicating that the supervised gating mechanism distributes documents across a finer partition of the embedding space.
Despite this broader routing breadth, the inverse relationship between routing load and lexical richness holds without exception, and the dominant expert maintains metrics close to the top-1000 baseline while minority experts consistently achieve the highest $H$, $R_{Voc}$, and $CP$ scores.
The absolute metric profiles differ across datasets (HotpotQA: $L_S \approx 15$--$20$, NQ: $L_S \approx 25$--$30$, MSMARCO: $L_S \approx 13$--$16$), yet the relative specialization pattern holds across both MoE variants and all three corpora.
These findings indicate that expert specialization is a property of the gating mechanism, with the intrinsic linguistic complexity of documents shaping how expert routing partitions the embedding space.

\begin{figure}[t]
    \centering
    
    \begin{subfigure}{\columnwidth}
        \centering
        \includegraphics[width=0.8\linewidth]{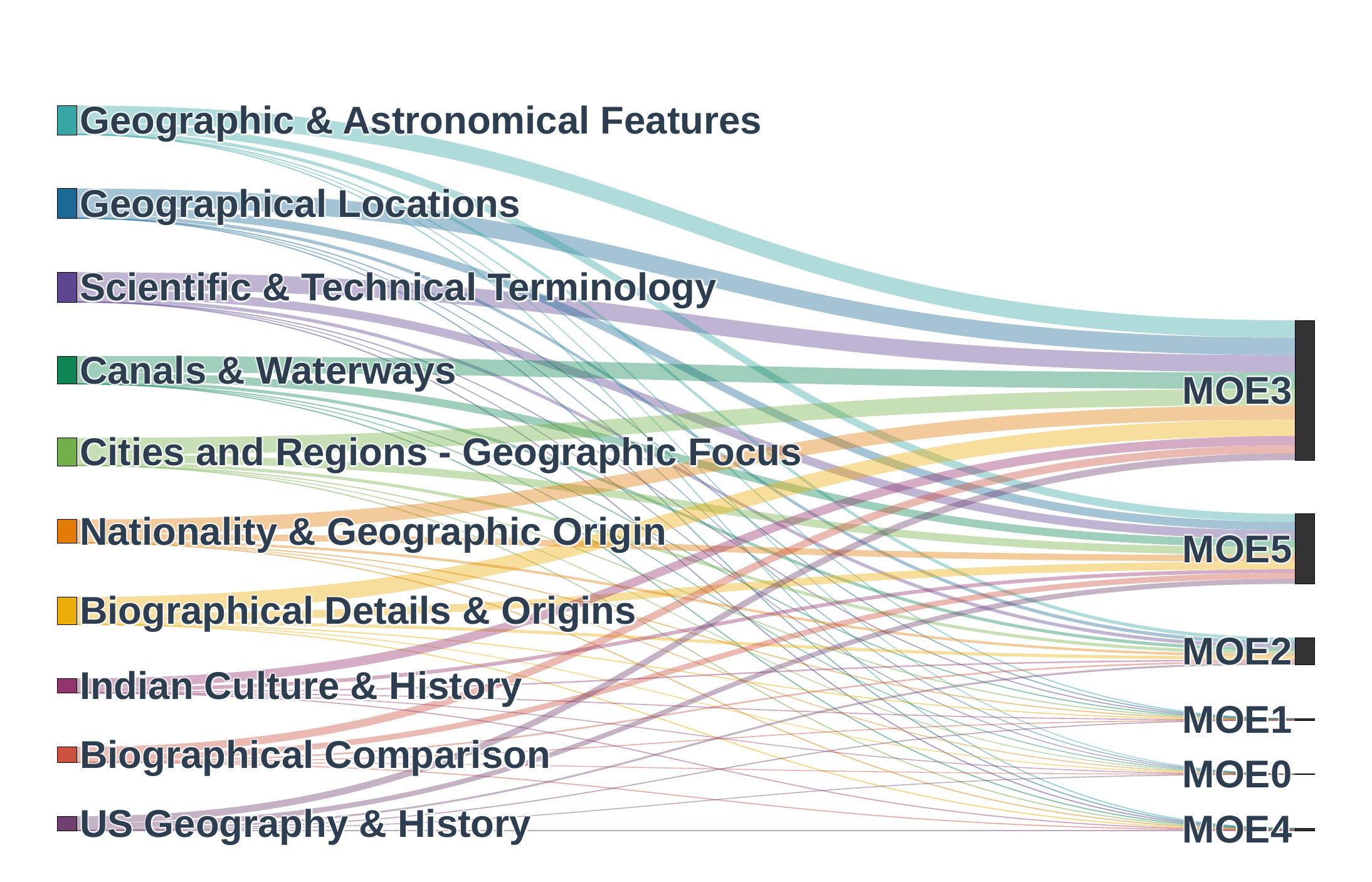}
        \caption{\textsc{Supervised} MoE for HotpotQA}
        \label{fig:sankey_sup}
    \end{subfigure}
    
    \vspace{0.5em}
    
    \begin{subfigure}{\columnwidth}
        \centering
        \includegraphics[width=0.8\linewidth]{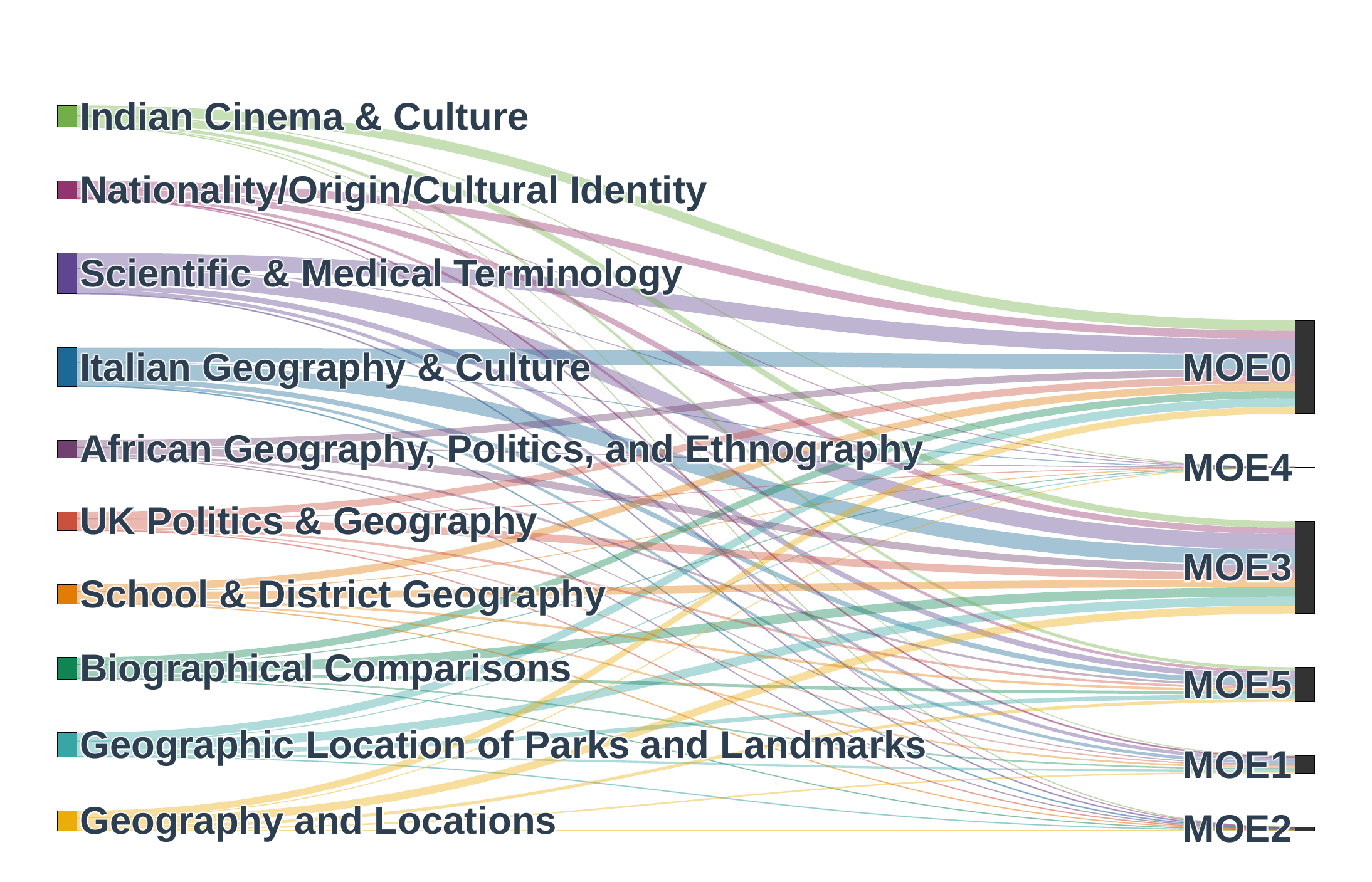}
        \caption{\textsc{Unsupervised} MoE for HotpotQA}
        \label{fig:sankey_unsup}
    \end{subfigure}
    
    \caption{Sankey diagrams displaying how concepts propagate through the experts.}
    \label{fig:sankey_concepts}
\end{figure}
\begin{figure}[t]
\centering
\includegraphics[width=0.7\linewidth]{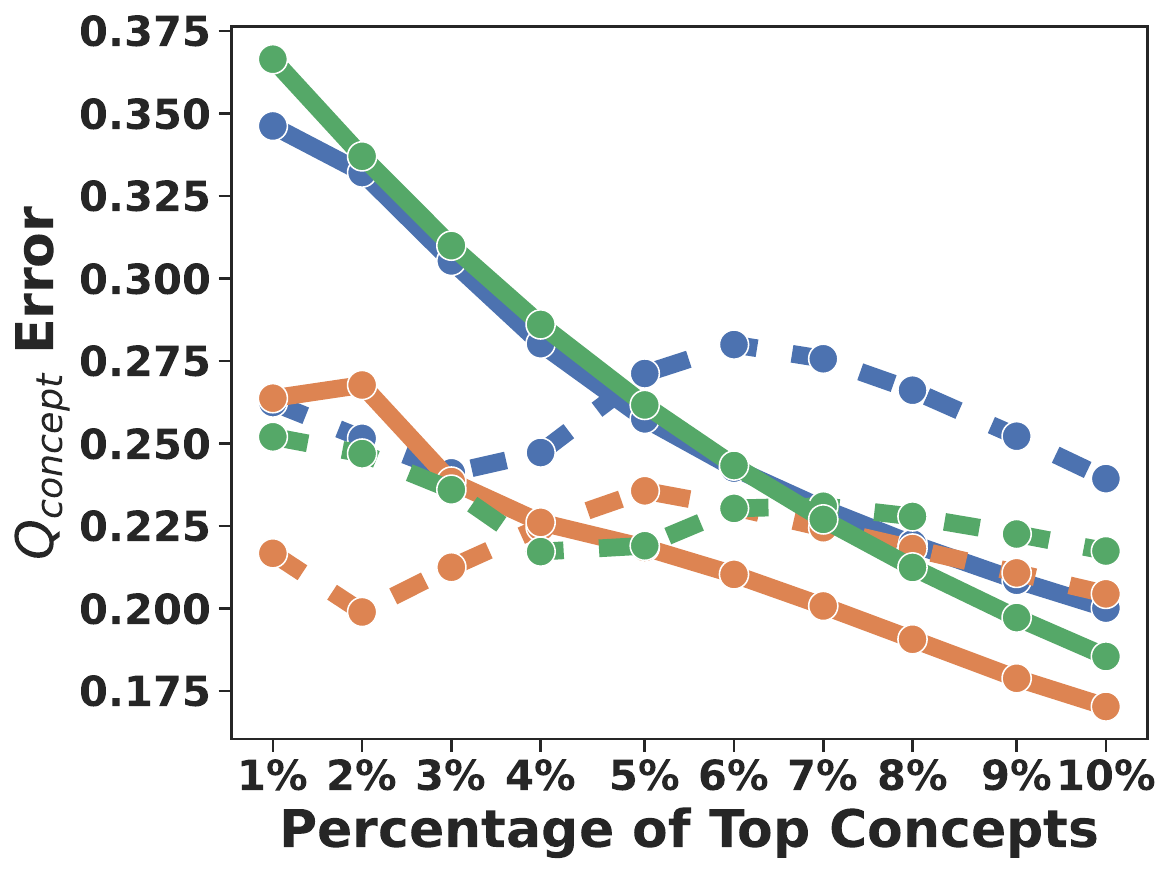}

\resizebox{0.65\columnwidth}{!}{%
\begin{tikzpicture}
\node[draw, fill=white, inner sep=1pt] {
\begin{tabular}{ccc}

\textcolor{blue}{$\bullet$} MSMarco &
\textcolor{orange}{$\bullet$} HotpotQA &
\textcolor{green!60!black}{$\bullet$} NQ \\[1pt]

\rule{8pt}{1.5pt} \textsc{Supervised} &
\raisebox{0.4ex}{\tikz \draw[dashed, line width=1pt] (0,0)--(0.5,0);} \textsc{Unsupervised} &
\\
\end{tabular}
};
\end{tikzpicture}
}%
\caption{We plot the $Q_\text{concept}$ error, including 1\%-10\% of the top concepts for each dataset (denoted by color) and embedding type (denoted by line type). }
\label{fig:concept_plot_error}
\end{figure}
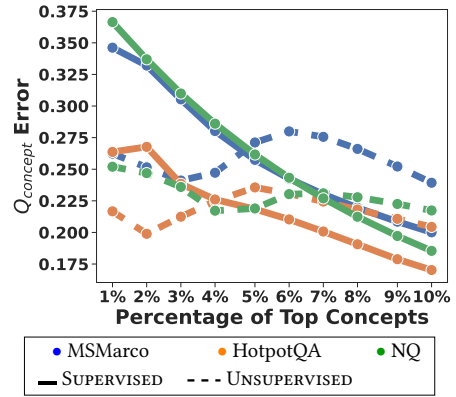
\subsection{RQ3: Concept Extraction and Expert Attribution}
Figure~\ref{fig:sankey_concepts} presents Sankey diagrams illustrating how the top-10 semantic concepts propagate across experts for both MoE variants.
In the \textsc{Supervised} embedding space, semantic groups formed by individual concepts cannot be localized to a single expert, indicating that experts do not 
specialize in isolated semantic topics.
As shown in Figure~\ref{fig:sankey_sup}, the dominant expert accounts for approximately two-thirds of the positive activations across nearly all concepts, aligning with the expert routing behavior observed in RQ2 and with the gating mechanism prediction distributions (Figure~\ref{fig:distributions}) reported by~\citet{DBLP:conf/emnlp/SokliPKP25}.
A similar specialization pattern occurs in the \textsc{Unsupervised} case, albeit with a different expert assignment distribution. 
Figure~\ref{fig:sankey_unsup} shows two primary experts sharing most of the concept activations, while activation levels are significantly lower for the remaining experts.
This observation repeated itself across both MoE variants across all datasets.
The broad distribution of concept activations across experts confirms that expert routing is not driven by semantic topic isolation.
We hypothesize that experts do not specialize in single concepts because of the intrinsic complexity of the documents, since a single document is typically characterized by the presence of multiple semantic themes.
Consequently, the gating mechanism likely relies on a broader set of characteristics for expert assignment, consistent with the linguistic specialization findings of RQ2.
To test this hypothesis, we analyzed the $Q_{concept}$ error, as depicted in Figure~\ref{fig:concept_plot_error}.
For the \textsc{supervised} variant, the error is lower when evaluating a smaller number of concepts. 
Conversely, in the \textsc{unsupervised} case, a smaller number of concepts induces a higher error. 
The differences in error between the two variants hover around 0.11 for NQ, 0.08 for MSMARCO, and 0.05 for HotpotQA. 
In the \textsc{supervised} case, when only 1\% (or 6) of the concepts are included in the set, neighboring documents share over two-thirds of them on average. 
This behavior remains relatively consistent regardless of the percentage of concepts included. 
Interestingly, the $Q_{concept}$ error for the \textsc{unsupervised} variant drops as the number of included concepts increases. 
This suggests that a broader set of concepts better explains neighborhood arrangements in the \textsc{unsupervised} embedding space. 
Notably, at the 10\% threshold, the \textsc{unsupervised} variant demonstrates a 0.04 to 0.05 improvement over its \textsc{supervised} counterpart. 
Ultimately, this analysis reveals that document neighborhoods are shaped by underlying data complexities that a single, monosemantic concept cannot adequately capture. 
This reinforces our conclusion that individual experts do not handle isolated concepts, as the relationships between documents are more complex than a single concept can explain.

\section{Conclusions} \label{sec:conclusions}
In this work, we introduce ExpertLens, a post-hoc explainability framework for Mixture-of-Experts (MoE)-enhanced dense retrievers that operates at the representation level, complementing existing feature-level attribution Explainability methods in Information Retrieval (ExIR).
ExpertLens leverages discriminative embedding space visualizations via DeepView (a discriminative variant of UMAP), for expert subspace characterization, and concept-level attribution to reveal how expert routing shapes the geometry of the embedding space.
Our empirical evaluation across five IR benchmarks and two MoE-enhanced dense retrievers with contrasting gating mechanisms shows that MoE routing improves the relative neighborhood structure around queries in the embedding space, while expert-induced subspace partitioning increases the overall geometric spread of the embedding space.
Expert specialization exhibits consistent linguistic profiles across all datasets, with general-purpose dominant experts absorbing structurally average documents, and minority experts forming lexically rich and concentrated subspaces.
Concept activations are broadly distributed across experts, with dominant experts accounting for most activations in all settings.
Document neighborhoods are shaped by a broad set of underlying semantic concepts, suggesting that expert assignment is driven by the linguistic properties of documents rather than by semantic isolation.
Together, these findings provide an interpretable view of how MoE routing shapes the embedding space of dense retrievers, linking embedding space geometry and expert specialization across IR tasks.

\appendix

\section{Prompt for Concept Labeling}\label{app:prompt}
In Figure \ref{fig:gemma_prompt}, we provide the prompt given to the Gemma-3-27b model to label the discovered concepts.

\begin{figure}[t]
    \centering
    \begin{tcolorbox}[
        width=\columnwidth,
        colback=gray!5!white,      
        colframe=gray!60!black,    
        title=\textbf{Prompt: Gemma-3-27b Concept Labeling}, 
        arc=2mm,                   
        boxrule=0.5pt,             
        fonttitle=\sffamily\small, 
        fontupper=\small           
    ]
    You are an expert in Semantic Analysis.
    I am analyzing a Sparse Autoencoder trained on search queries.
    Below are several groups of queries. Each group activates a specific latent feature.
    
    Your task is to analyze each group and identify the shared intent, topic, or linguistic feature.
    Analyze the following groups of search queries and assign a semantic label.
    
    \vspace{0.5em}
    \textbf{INSTRUCTIONS:}
    \begin{itemize}
        \setlength{\itemsep}{0pt}
        \setlength{\parskip}{0pt}
        \item Return a valid JSON object.
        \item Keys must be the "Group ID" strings.
        \item Do NOT add any markdown formatting, explanation, or chatter. 
        \item Start immediately with \{ and end with \}.
    \end{itemize}
    
    \textbf{REQUIRED JSON FORMAT:}
\begin{verbatim}
{
  "182": {
    "label": "Crime Films"
  },
  "196": {
    "label": "Literary and Cultural References",
    "description": "Queries about authors, books,
                    comics, dolls, and magazines."
  },
  "250": {
    "label": "Linguistics",
    "description": "Queries about languages, words,
                    and linguistic concepts."
  },
}
\end{verbatim}
    \end{tcolorbox}
    \caption{The zero-shot prompt provided to Gemma-3-27b to automatically generate semantic labels and descriptions for discovered SAE concepts based on top-activating queries. The shown concepts are from the HotpotQA dataset.}
    \label{fig:gemma_prompt}
\end{figure}

\begin{acks}
This work has received funding from the European Union’s Horizon Europe research and innovation programme under the Marie Skłodowska-Curie grant agreement No 101073307.
\end{acks}

\bibliographystyle{ACM-Reference-Format}
\bibliography{refs}










\end{document}

%% file: phase1.tex
\begin{table*}[t]
\caption{
Assessment of the neighborhood structure within three different embedding spaces, as these are formulated with and without the employment of two contrastive MoE variants across all datasets. 
Scores measure the precision@k per model and dataset at different cutoffs (\%), as well as the absolute proximity of each relevant document $d^+$ to its corresponding query $q$ via cosine distance. 
Best results for each setting are in \textbf{bold} text.
}
\label{tab:results_knn}
\centering
\resizebox{1\linewidth}{!}{%
\begin{tabular}{ll|lll|lll|lll|lll|lll}
\multicolumn{2}{r|}{}       & \multicolumn{3}{c|}{TREC19} & \multicolumn{3}{c|}{TREC20} & \multicolumn{3}{c|}{MSMARCO} & \multicolumn{3}{c|}{NQ} & \multicolumn{3}{c}{HotpotQA} \\
Retriever & Variant & \multicolumn{1}{l}{{\textbf{P@20}}} & \multicolumn{1}{l}{{\textbf{P@100}}} & \multicolumn{1}{l|}{{\textbf{$\text{cos}_{dist}\downarrow$}}} 
                    & \multicolumn{1}{l}{{\textbf{P@20}}} & \multicolumn{1}{l}{{\textbf{P@100}}} & \multicolumn{1}{l|}{{\textbf{$\text{cos}_{dist}\downarrow$}}}
                    & \multicolumn{1}{l}{{\textbf{P@1}}} & \multicolumn{1}{l}{{\textbf{P@2}}} & \multicolumn{1}{l|}{{\textbf{$\text{cos}_{dist}\downarrow$}}}
                    & \multicolumn{1}{l}{{\textbf{P@1}}} & \multicolumn{1}{l}{{\textbf{P@2}}} & \multicolumn{1}{l|}{{\textbf{$\text{cos}_{dist}\downarrow$}}}
                    & \multicolumn{1}{l}{{\textbf{P@1}}} & \multicolumn{1}{l}{{\textbf{P@2}}} & \multicolumn{1}{l}{{\textbf{$\text{cos}_{dist}\downarrow$}}} \\ \hline

\multirow{3}{*}{TinyBERT}   
& \textsc{fine-tuned}
&  .333 & .171 & \textbf{.341} & .276 & .107 & \textbf{.334} & .120 & .095 & \textbf{.243} & .129 & .105 & \textbf{.239} & .256 & .175 & \textbf{.349} \\
 & \textsc{unsupervised} 
& \textbf{.341} & \textbf{.175} & .368 & .276 & .107 & .360 & .118 & .094 & .262 & .137 & .113 & .268 & \textbf{.263} & \textbf{.178} & .382 \\
 & \textsc{supervised}  
& .330 & .171 & .417 & \textbf{.278} & \textbf{.109} & .406 & \textbf{.123} & \textbf{.095} & .298 & \textbf{.148} & \textbf{.118} & .309 & .260 & .177 & .406 \\ \hline

\multirow{3}{*}{BERT-Small} 
& \textsc{fine-tuned}
& .344 & .184 & \textbf{.193} & .268 & \textbf{.114} & \textbf{.191} & .115 & .097 & \textbf{.157} & .085 & .075 & \textbf{.202} & .308 & .217 & .263 \\
 & \textsc{unsupervised} 
& \textbf{.349} & \textbf{.185} & .220 & .268 & .113 & .218 & \textbf{.115} & \textbf{.097} & .180 & .088 & .078 & .225 & .316 & .222 & \textbf{.259} \\
 & \textsc{supervised}   
& .3221& .181 & .277 & \textbf{.274} & .112 & .275 & .109 & .094 & .227 & \textbf{.097} & \textbf{.082} & .261 & \textbf{.341} & \textbf{.240} & .286 \\ \hline

\multirow{3}{*}{BERT-Base}  
& \textsc{fine-tuned}
& \textbf{.400} & .203 & \textbf{.266} & .319 & .130 & \textbf{.267} & .134 & \textbf{.112} & \textbf{.206} & .172 & .142 & \textbf{.210} & .467 & .324 & \textbf{.253} \\
 & \textsc{unsupervised} 
& .385 & .201 & .267 & \textbf{.330} & .130 & .268 & \textbf{.135} & .108 & .207 & .173 & .142 & .213 & .450 & .315 & .258 \\
 & \textsc{supervised}   
& .387 & \textbf{.205} & .297 & .328 & \textbf{.132} & .299 & .129 & .109 & .229 & \textbf{.194} & \textbf{.158} & .228 & \textbf{.491} & \textbf{.337} & .288 \\ \hline

\multirow{3}{*}{ColBERT} 
& \textsc{fine-tuned}
& .433 & .226 & \textbf{.262} & .354 & .139 & \textbf{.256} & .143 & .120 & \textbf{.192} & .200 & .165 & \textbf{.198} & .440 & .304 & \textbf{.255} \\
 & \textsc{unsupervised} 
& .429 & .226 & .267 & \textbf{.357} & \textbf{.142} & .262 & \textbf{.147} & \textbf{.122} & .195 & .205 & .167 & .204 & .453 & .314 & .275 \\
 & \textsc{supervised}   
& \textbf{.452} & \textbf{.226} & .284 & .349 & .139 & .284 & .141 & .118 & .207 & \textbf{.213} & \textbf{.177} & .214 & \textbf{.466} & \textbf{.320} & .297  
\end{tabular}
}
\end{table*}

%% file: phase2_unsup.tex
\begin{table*}[t]
\centering
\scriptsize
\renewcommand{\arraystretch}{1}
\setlength{\tabcolsep}{2.8pt}
\caption{Textual characteristics analysis for \textsc{Unsupervised} (SB-MoE),
reporting the mean of each metric across the top-100 DBSCAN-selected queries per dataset.
Considering active experts only, i.e., those handling at least 10 documents for a given query on average per dataset.
\textbf{Bold}: highest value per expert per dataset;
\underline{underline}: lowest value per expert per dataset.}
\label{tab:phase2_unsup}
\par\smallskip
\resizebox{\linewidth}{!}{%
\begin{tabular}{@{}
  l
  r r rrr
  @{\hspace{6pt}}
  r r rrrr
  @{\hspace{6pt}}
  r r rrrrr
  @{}}
\toprule
& \multicolumn{5}{c}{\textit{MSMARCO}}
& \multicolumn{6}{c}{\textit{NQ}}
& \multicolumn{7}{c}{\textit{HotpotQA}} \\
\cmidrule(lr){2-6}\cmidrule(lr){7-12}\cmidrule(lr){13-19}
\textbf{Metric}
  & \textbf{Corpus} & \textbf{top-1000} & \textbf{E0} & \textbf{E3} & \textbf{E4}
  & \textbf{Corpus} & \textbf{top-1000} & \textbf{E0} & \textbf{E2} & \textbf{E4} & \textbf{E5}
  & \textbf{Corpus} & \textbf{top-1000} & \textbf{E0} & \textbf{E1} & \textbf{E2} & \textbf{E3} & \textbf{E5} \\
\midrule
Docs
  & 8.8M & 1000 & 477.23 & \underline{38.35} & \textbf{480.64}
  & 2.6M & 1000 & \textbf{732.10} & 27.38 & \underline{15.94} & 217.39
  & 5.2M & 1000 & \textbf{418.99} & 85.59 & \underline{13.24} & 367.33 & 114.24 \\
$ari\downarrow$
  & $8.762$ & $8.500$
  & $\underline{7.742}$
  & $8.746$
  & ${9.234}$
  & $14.329$ & $13.521$
  & ${13.653}$
  & $12.618$
  & $\underline{11.485}$
  & $12.975$
  & $10.270$ & $10.115$
  & $\underline{10.004}$
  & $10.270$
  & ${10.947}$
  & $10.143$
  & $10.367$ \\
$H$
  & $0.542$ & $0.765$
  & $\underline{0.792}$
  & $\mathbf{0.875}$
  & $0.796$
  & $0.562$ & $0.751$
  & $\underline{0.761}$
  & $0.889$
  & $\mathbf{0.902}$
  & $0.813$
  & $0.525$ & $0.736$
  & $\underline{0.771}$
  & $0.852$
  & $\mathbf{0.907}$
  & $0.776$
  & $0.838$ \\
$R_{\text{Voc}}$
  & $0.003$ & $0.182$
  & $\underline{0.262}$
  & $\mathbf{0.549}$
  & $0.264$
  & $0.006$ & $0.180$
  & $\underline{0.208}$
  & $0.663$
  & $\mathbf{0.731}$
  & $0.373$
  & $0.013$ & $0.205$
  & $\underline{0.295}$
  & $0.525$
  & $\mathbf{0.732}$
  & $0.299$
  & $0.476$ \\
$CP$
  & $0.600$ & $0.674$
  & $\underline{0.703}$
  & $\mathbf{0.798}$
  & $0.707$
  & $0.583$ & $0.649$
  & $\underline{0.659}$
  & $0.810$
  & $\mathbf{0.843}$
  & $0.714$
  & $0.656$ & $0.695$
  & $0.723$
  & $0.783$
  & $\mathbf{0.829}$
  & $\underline{0.721}$
  & $0.767$ \\
$L_S$
  & $15.635$ & $14.873$
  & $\underline{14.399}$
  & $\mathbf{15.892}$
  & $15.260$
  & $25.149$ & $24.519$
  & $\mathbf{24.777}$
  & $22.567$
  & $\underline{19.732}$
  & $23.487$
  & $16.770$ & $17.202$
  & $\underline{17.031}$
  & $17.208$
  & $\mathbf{17.996}$
  & $17.310$
  & $17.354$ \\
\bottomrule
\end{tabular}
}%
\end{table*}

%% file: phase2_sup.tex
\begin{table*}[t]
\centering
\scriptsize
\renewcommand{\arraystretch}{1}
\setlength{\tabcolsep}{2.8pt}
\caption{Textual characteristics analysis for \textsc{Supervised} (DenseC3),
reporting the mean of each metric across the top-100 DBSCAN-selected queries per dataset.
Considering active experts only, i.e., those handling at least 10 documents for a given query on average per dataset.
\textbf{Bold}: highest value per expert per dataset;
\underline{underline}: lowest value per expert per dataset.}
\label{tab:phase2_sup}
\par\smallskip
\resizebox{\linewidth}{!}{%
\begin{tabular}{@{}
  l
  r r rrrrr
  @{\hspace{6pt}}
  r r rrrrr
  @{\hspace{6pt}}
  r r rrr
  @{}}
\toprule
& \multicolumn{7}{c}{\textit{MSMARCO}}
& \multicolumn{7}{c}{\textit{NQ}}
& \multicolumn{5}{c}{\textit{HotpotQA}} \\
\cmidrule(lr){2-8}\cmidrule(lr){9-15}\cmidrule(lr){16-20}
\textbf{Metric}
  & \textbf{Corpus} & \textbf{Top-1k}
  & \textbf{E1} & \textbf{E2} & \textbf{E3} & \textbf{E4} & \textbf{E5}
  & \textbf{Corpus} & \textbf{Top-1k}
  & \textbf{E1} & \textbf{E2} & \textbf{E3} & \textbf{E4} & \textbf{E5}
  & \textbf{Corpus} & \textbf{Top-1k}
  & \textbf{E2} & \textbf{E3} & \textbf{E5} \\
\midrule
Docs
  & 8.8M & 1000 & 96.07 & 146.22 & \textbf{484.94} & \underline{51.96} & 212.27
  & 2.6M & 1000 & 70.25 & 93.58 & \textbf{528.87} & \underline{34.09} & 268.46
  & 5.2M & 1000 & \underline{79.10} & 451.40 & \textbf{451.86} \\
$ari\downarrow$
  & $8.762$ & $8.839$
  & $\underline{}{8.337}$
  & ${9.471}$
  & $8.505$
  & $9.347$
  & $9.398$
  & $14.329$ & $13.051$
  & $12.904$
  & $\underline{}{12.358}$
  & ${13.573}$
  & $14.819$
  & $12.583$
  & $10.270$ & $9.993$
  & ${11.023}$
  & $\underline{9.252}$
  & $10.508$ \\
$H$
  & $0.542$ & $0.756$
  & $0.839$
  & $0.819$
  & $\underline{0.778}$
  & $\mathbf{0.860}$
  & $0.814$
  & $0.562$ & $0.749$
  & $0.840$
  & $0.832$
  & $\underline{0.772}$
  & $\mathbf{0.860}$
  & $0.789$
  & $0.525$ & $0.734$
  & $\mathbf{0.833}$
  & $\underline{0.762}$
  & $0.767$ \\
$R_{\text{Voc}}$
  & $0.003$ & $0.163$
  & $0.383$
  & $0.338$
  & $\underline{0.223}$
  & $\mathbf{0.470}$
  & $0.325$
  & $0.006$ & $0.179$
  & $0.483$
  & $0.456$
  & $\underline{0.248}$
  & $\mathbf{0.553}$
  & $0.294$
  & $0.013$ & $0.197$
  & $\mathbf{0.468}$
  & $0.282$
  & $\underline{0.269}$ \\
$CP$
  & $0.600$ & $0.678$
  & $0.754$
  & $0.731$
  & $\underline{0.710}$
  & $\mathbf{0.776}$
  & $0.721$
  & $0.583$ & $0.648$
  & $0.744$
  & $0.738$
  & $\underline{0.678}$
  & $\mathbf{0.766}$
  & $0.683$
  & $0.656$ & $0.697$
  & $\mathbf{0.760}$
  & $0.739$
  & $\underline{0.708}$ \\
$L_S$
  & $15.635$ & $15.352$
  & $\underline{14.439}$
  & $\mathbf{16.787}$
  & $14.506$
  & $16.733$
  & $16.541$
  & $25.149$ & $24.569$
  & $24.890$
  & $\underline{22.973}$
  & $25.398$
  & $\mathbf{27.495}$
  & $23.927$
  & $16.770$ & $16.951$
  & $\mathbf{18.978}$
  & $\underline{15.643}$
  & $17.893$ \\
\bottomrule
\end{tabular}
}%
\end{table*}